\documentclass{article}

\usepackage{graphicx}

\def\ligne#1 {\hbox to\hsize{#1}}
\def\leurre{\noindent\leftskip 0pt\footnotesize\baselineskip 10pt\parindent 0pt}

\newtheorem{fig}{{\sc Figure}}

\newtheorem{tab}{{\sc Table}} 

\newtheorem{thm}{{\sc Theorem}}

\begin{document}
\ligne{\hfill
\bf\Large A weakly universal cellular automaton in the\hfill} 
\ligne{\hfill \bf\Large heptagrid with three states\hfill}
\vskip 20pt

\ligne{\hfill Maurice {\sc Margenstern}\hfill}
\vskip 30pt
\ligne{\hfill Universit\'e de Lorraine\hfill}
\ligne{\hfill LITA, EA 3097,\hfill} 
\ligne{\hfill Campus du Saulcy,\hfill}
\ligne{\hfill 57045 Metz, C\'edex 1, France\hfill}
\ligne{\hfill {\it e-mail}: {\tt maurice.margenstern@univ-lorraine.fr, margenstern@gmail.com}
\hfill}
\vskip 20pt
{\bf Abstract} {\it\small
In this paper, we construct a cellular automaton on the heptagrid which is planar,
weakly universal and which have three states only. This result improves the best result
which was with four states.}


\vskip 10pt



\section{Introduction}

   In this paper, we construct a weakly universal cellular automaton on the heptagrid,
see Theorem~\ref{univ3} at the end of the paper. Two papers, \cite{mmsyENTCS,mmhepta4}
already constructed such a cellular automaton, the first one with 6~states, the second one with
4~states. In this paper, the cellular automaton we construct has three states only.
It uses the same principle of simulating a register machine through a railway circuit, but
the implementation takes advantage of new ingredients introduced by the author in his quest
to lower down the number of states, see~\cite{mmbook3}. It also introduce new constructions
which significantly improve the scenario used in the previous papers.
The reader is referred to~\cite{mmbook1,mmbook2,mmbook3} for an introduction to hyperbolic
geometry turned to the implementation of cellular automata in this context. A short introduction
can also be found in~\cite{mmDMTCS}. However, it is not required to be an expert in hyperbolic 
geometry in order to read this paper.

We refer the reader to~\cite{mmMCUZu} for a discussion
about weak universality. Here, we just remember the main characteristics of our cellular
automaton with respect to universality.  The simulation is performed by a locomotive
running on a railway circuit, using the basic elements introduced in~\cite{stewart}.
The simulation is a planar one in this meaning that
the trajectory of the locomotive involves infinitely many cycles when the computation does not
halt. Now, the initial configuration is infinite but, outside a finite domain, it consists
of two periodic structures. This is why the automaton is called weakly universal.
In the paper, we remember the basic model we use in Section~\ref{railway}
and we stress on the new features in Section~\ref{implement} were we thoroughly describe
the implementation of the model in the heptagrid. In Section~\ref{inhepta}, we remind the
reader with the general setting of the heptarid, the tiling~$\{7,3\}$ of the hyperbolic plane.
Section~\ref{rules} gives the rule of the cellular automaton, which allows us to prove
Theoreme~\ref{univ3}.

\section{Basic features}
\label{railway}

   The railway circuit, mentioned in the introduction consists of tracks, crossings and 
switches, and a single locomotive
runs over the circuit. The tracks are pieces
of straight line or arcs of a circle. In the hyperbolic context, we shall replace these
features by assuming that the tracks travel either on {\bf verticals} or {\bf horizontals}
and we shall make it clear a bit later what we call by these words. The crossing is
an intersection of two tracks, and the locomotive which arrives at an intersection 
by following a track goes on by the track which naturally continues the track through which
it arrived. Again, later we shall make it clear what this natural continuation is.
Below, Figure~\ref{switches} illustrates the switches and Figure~\ref{element} illustrates
the use of the switches in order to implement a memory element which exactly contains 
one bit of information.

   The three kinds of switches are the {\bf fixed switch}, the {\bf flip-flop}
and the {\bf memory switch}. In order to understand how the switches work, notice that in
all cases, three tracks abut the same point, the {\bf centre} of the switch. On one side
switch, there is one track, say~$a$, and on the other side, there are two tracks, call 
them~$b$ and~$c$. When the locomotive arrive through~$a$, we say that it is an {\bf active}
crossing of the switch. When it arrives either through~$b$ or~$c$, we say that it is a
{\bf passive} crossing. 

\vtop{
\vspace{5pt}
\ligne{\hfill
\includegraphics[scale=1.0]{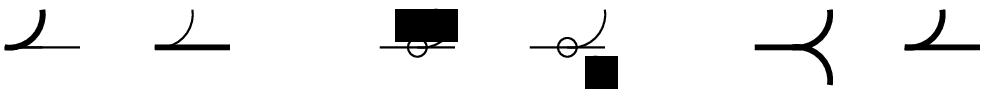}
\hfill}
\vspace{-5pt}
\begin{fig}\label{switches}
\leurre
The switches of the railway circuit. From left to right: the fixed switch, the flip-flop
and the memory switch.
\end{fig}
\vspace{10pt}
}

   In the fixed switch, in an active passage, the locomotive is sent either always to~$b$
or always to~$c$, we say that the {\bf selected track} is always~$b$ or it is always~$c$. 
In the passive crossing, the switch does nothing, the locomotive leaves
the switch through~$a$. In the flip-flop, passive crossings are prohibited: the circuit
must be managed in such a way that a passive crossing never occurs at any flip-flop.
During an active passage, the selected track is changed just after the passage of the locomotive:
if it was~$b$, $c$ before the crossing, it becomes~$c$, $b$ respectively after it.
In the memory switch, both active and passive crossings are allowed. The selected tracks also
may change and the change is dictated by the following rule: after the first crossing, only
in case it is active, the selected track is always defined as the track taken by the locomotive
during its last passive crossing of the switch. The selected track at a given switch defines
its {\bf position}.

   The current configuration of the circuit is the position of all the switches of the circuit.
Note that it may be coded in a finite word, even if the circuit is infinite, as at each time,
only finitely many switches have been visited by the locomotive.

   Figure~\ref{element} illustrates how a flip-flop and a memory switch can be coupled 
in order to make a one bit memory element.

\vtop{
\vspace{-5pt}
\ligne{\hfill
\includegraphics[scale=0.81]{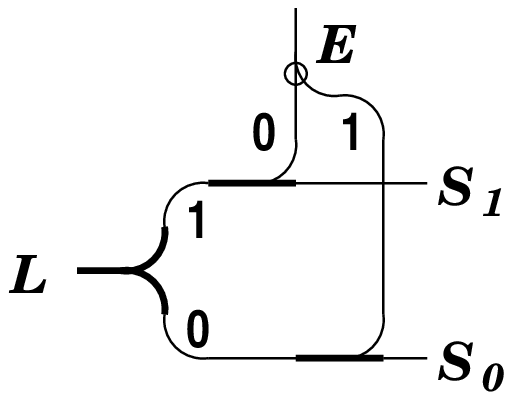}
\hfill}
\ligne{\hfill
\includegraphics[scale=0.54]{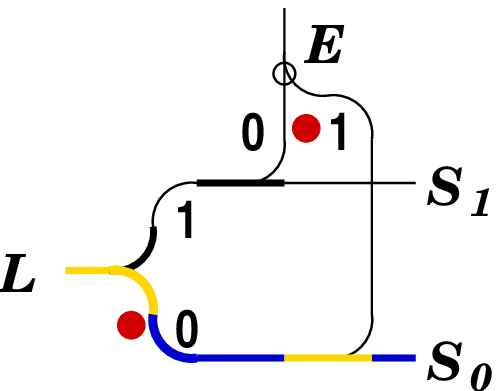}
\includegraphics[scale=0.54]{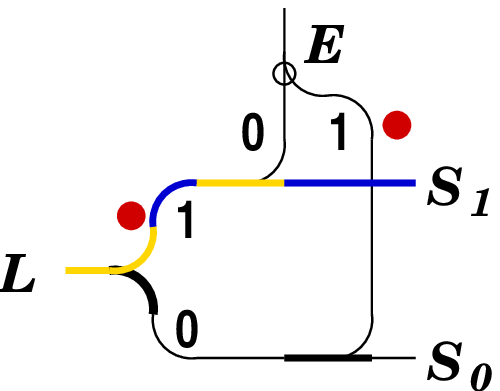}
\hfill}
\ligne{\hfill
\includegraphics[scale=0.45]{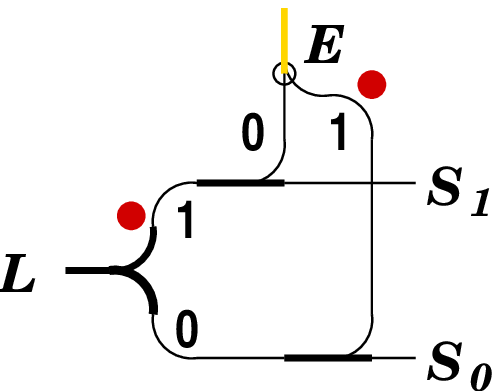}
\includegraphics[scale=0.45]{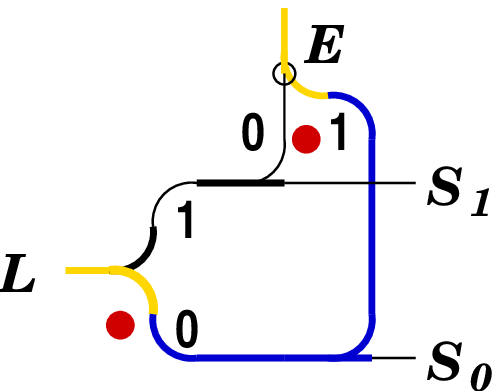}
\includegraphics[scale=0.45]{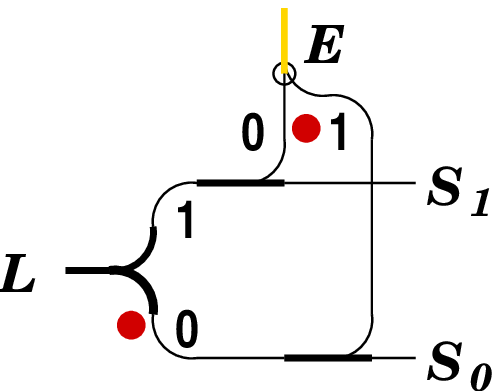}
\includegraphics[scale=0.45]{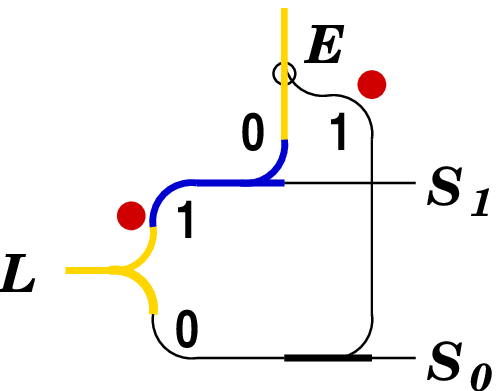}
\hfill}
\vspace{-5pt}
\begin{fig}\label{element}
\leurre
The basic element of the circuit.
\vskip 0pt
Second row: reading the element, $0$, then~$1$. Third row: first
two elements: writing~$0$; last two elements: writing~$1$.
\end{fig}
}

\subsection{In the heptagrid}
\label{inhepta}

   As mentioned in the introduction, the first weakly universal cellular automaton on the
heptagrid was done by the author and a co-author, see~\cite{mmsyENTCS}. The paper implements
the solution sketchily mentioned in Subsection~\ref{railway} with 6 states. In the next paper 
about a weakly universal cellular automaton on the heptagrid, see~\cite{mmhepta4}, the same model
is implemented with 4~states. In~\cite{mmsyENTCS} 
the cells of the tracks where the locomotive runs have a specific colour and the locomotive
is implemented with two cells, the front and the rear. The front is green and the rear is red
while the tracks are blue. Later, in~\cite{mmhepta4}, the tacks are changed: the cells where
the locomotive runs are also in the quiescent state when the locomotive
is far away. The tracks are delimited by milestones which are regularly put along the track.
The milestones are most often blue. The locomotive is still implemented with two cells,
the front being blue, as the milestones, and the rear being red.
As in many of the indicated papers, the centre of the switch is signalized by the neighbouring 
of the centre.

\subsubsection{Former implementations}

   Figure~\ref{hca73stables} illustrates the implementation of the crossing and of the
switches performed in~\cite{mmsyENTCS}, showing in particular, the feature at which we just 
pointed. 
Figure~\ref{hcaelement} shows a global view of how the tree structure of the tiling can be used 
to implement a basic element in the heptagrid.
%
%
We have the following results.

\begin{thm}\label{univhepta1} {\rm (Margenstern, Song),} {\it cf.}{\rm\cite{mmsyENTCS}}
$-$ There is a planar cellular automaton on the heptagrid with $6$-states
which is weakly universal and rotation invariant.
\end{thm}

\def\WW{\hbox{\tt W}{}}
\def\BB{\hbox{\tt B}{}}
\def\GG{\hbox{\tt G}{}}
\def\RR{\hbox{\tt R}{}}
\def\DB{\hbox{\tt B$_D$}{}}
\def\DG{\hbox{\tt G$_D$}{}}
\def\LG{\hbox{\tt G$_L$}{}}
\def\DR{\hbox{\tt R$_D$}{}}
\def\XX{\hbox{\tt X}{}}

\begin{thm}\label{univhepta2} {\rm (Margenstern)} {\it cf.}{\rm\cite{mmhepta4}}
$-$ There is a planar cellular automaton on the heptagrid with $4$-states
which is weakly universal and rotation invariant.
\end{thm}

\vtop{
\vspace{-5pt}
\ligne{\hfill
\includegraphics[scale=0.8]{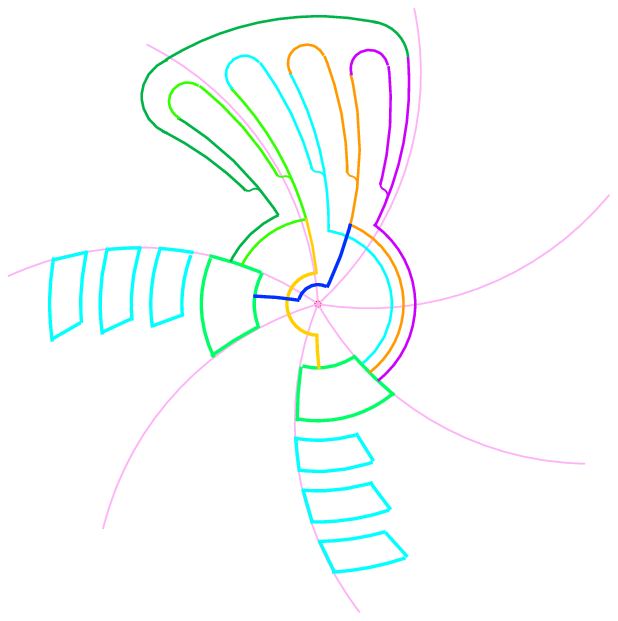}
\hfill}
\vspace{-20pt}
\begin{fig}\label{hca_univglobal}
\leurre
Global look of the implementation of a toy example in the heptagrid. Note the implementation
of the two registers: each one along a branch in a sector of the heptagrid.
\end{fig}
\vspace{10pt}
}


   Figure~\ref{hca_univglobal} illustrates the configuration of the railway circuit implementing
a toy program of a register machine. All the constructions performed in previous papers and in
this one too produce a configuration whose guidelines obey those illustrated by
Figure~\ref{hca_univglobal}.

   Now, figures~\ref{element} and~\ref{hcaelement} show that the implementation of the basic element
involves two cycles. As a register consists of infinitely many copies of a unit, again look at
Figure~\ref{hca_univglobal}, and as a unit contains
two basic elements, we can see that a non halting computation entails that the locomotive
runs over infinitely many cycles, so that the computation is planar.

   Figure~\ref{hca73stables} illustrates the idle configuration of the crossing and the switches
of the circuit for the cellular automaton constructed in the proof of Theorem~\ref{univhepta1}.
By idle configuration, we mean a configuration where the locomotive is not present.

\vtop{
\vspace{-25pt}
\ligne{\hskip -15pt
\includegraphics[scale=0.5]{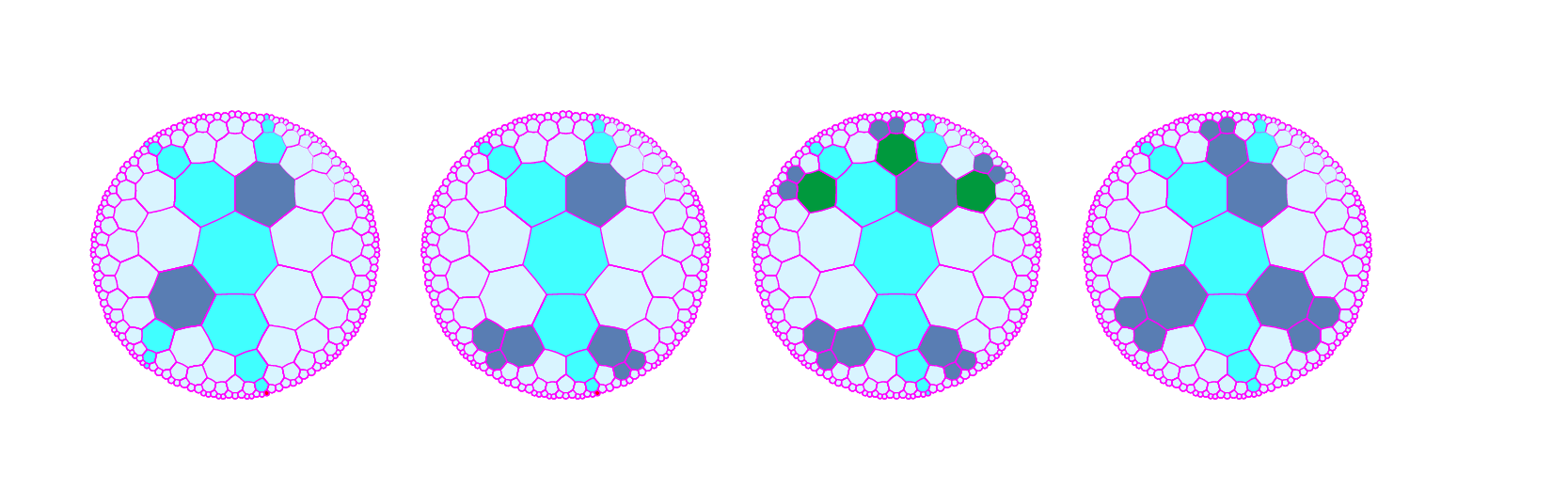}
\hfill}
\vspace{-20pt}
\begin{fig}\label{hca73stables}
\leurre
Implementation of the crossing and of the switches in the heptagrid for the 
weakly universal cellular automaton with $6$ states.
\vskip 0pt
From left to right: the crossing, the fixed switch, the flip-flop and the memory switch.
\end{fig}
\vspace{10pt}
}

\section{The scenario of our simulation}
\label{implement}

   Most of the cellular automata in hyperbolic spaces I constructed, myself or with
a co-author, apply the same model of computation. 
This general scenario is described in detail in~\cite{mmbook2,mmbook3}. Here,
we simply give the guidelines in order to introduce the changes which are specific to this
implementation.

\subsubsection{The new scenario}
\label{newscenar}

   In this paper, we take benefit of various improvements which I brought in the construction
of weakly universal cellular automaton constructed in other contexts: 
in the hyperbolic $3D$-space and in the tiling
$\{13,3\}$ of the hyperbolic plane, that latter automaton having two states only,
see~\cite{mmbook3} for details. But here we need something more, as soon explained.

   Our implementation follows the same general simulation as the one described in 
Subsection~\ref{railway}. In particular, Figure~\ref{hcaelement} is still meaningful in
this new setting.

\vtop{
\vspace{-5pt}
\ligne{\hfill
\includegraphics[scale=0.5]{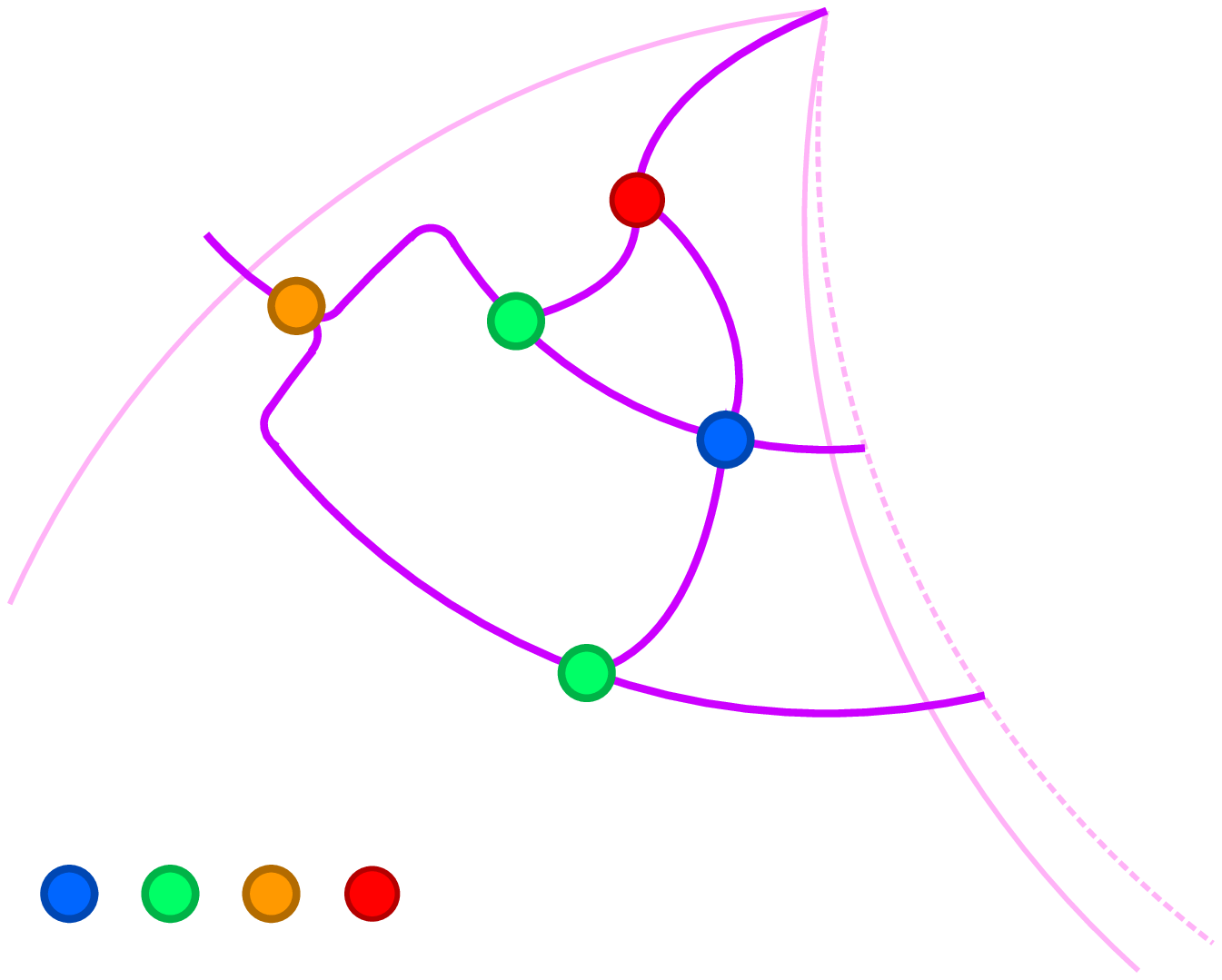}
\hfill}
\begin{fig}\label{hcaelement}
\leurre
Implementation of the basic element in the heptagrid.\\
The discs in the bottom line represent, from left to right: the crossing, the fixed switch,
the memory switch and the flip-flop.
\end{fig}
\vspace{10pt}
 }

   However, here, new features are introduced.

   The first change is that the tracks are one-way. In some sense this is closer to what
we can see for railways in real life, in particular for high-speed ones. This change entails
a big change in the switches and in the crossings. There is no change for the flip-flop which
was already a one-way structure from the very beginning as passive crossings are ruled out
for this kind of switches. For fixed switches it introduces a very small change: for a passive
crossing, we keep the
structure and for the active one, as the selected track is the same, it 
is enough to continue the active way without branching at the centre of the switch,
see Figure~\ref{newswitches}. In the same picture, we can see that the situation is different
for the memory switch. This time, as there are two possible crossings of the switch and as the 
selected track may change, we have two one-way switches: an active one and a passive one. 
At first glance, the active switch looks like a flip-flop and the passive switch looks like
a one-way fixed one. However, due to the working of the memory switch, we could say that the
active memory switch is passive while the passive memory switch is active. Indeed, 
during an active crossing, the selected track is not changed contrary to what happens in the
case of the flip-flop. Now, during a passive crossing, the switch looks at which track is
crossed: the selected or the non-selected one. If it is the non-selected one, then the selection
is changed and this change is also transferred to the active switch. Accordingly, there is
a connection between the active and the passive one-way memory switches.

\vtop{
\vspace{-5pt}
\ligne{\hfill
\includegraphics[scale=0.5]{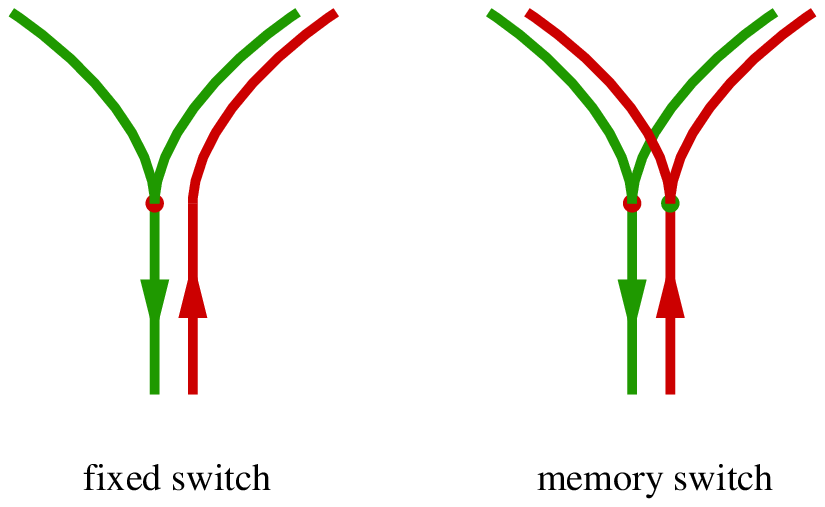}
\hfill}
\vspace{-20pt}
\begin{fig}\label{newswitches}
\leurre
The new switches for a one-way structured circuit: the fixed and the memory ones. Note that
the flip-flop remains the same as in Figure~{\rm\ref{switches}}.
\end{fig}
\vspace{10pt}
}

   Now, if we wish to significantly reduce the number of states, we also have to 
change the locomotive: it is now reduced to a single cell which, in principle is made
possible by the fact that the tracks are one-way.
This is a change with respect to 
the wealky universal cellular automaton with 4~states, see~\cite{mmhepta4}.
But this is not enough: we also have to change the crossings. Contrary to what happens
in the $3D$-space where crossings can be replaced by bridges, which makes the situation
significantly easier, crossings cannot be avoided in the plane. 

   In~\cite{mmbook3} we indicate a solution which allowed me to build a weakly universal
cellular automaton in the hyperbolic plane with 2~states only. However, this was not performed
in the pentagrid nor in the heptagrid, but in the tiling $\{13,3\}$. This solution was already 
implemented in the pentagrid, which allowed me to reduce the number of states 
from~9 down to~5. Here, we improve the result for four states downto 3 states.
However, the implementation of the new crossing is somehow tuned from what
is done in~\cite{mmbook3} and in~\cite{mmpenta5}: The pattern at the branching 
of~\cite{mmbook3,mmpenta5} is changed: a doubling structure is put on the track arriving
to the crossing. 

   First, we look at a crossing of two one-way tracks. The main idea is that we organize
the crossing in view of a {\bf round-about}: an interference of road trafic in our railway 
circuit. Figure~\ref{rondpointsimple} illustrates this structure which we call the {\bf simple
round-about}.
We may notice that the locomotive arriving either from~{\bf A} or~{\bf B} in
Figure~\ref{rondpointsimple} has to turn right at the second pattern it meets on its way.
Note that the locomotive arrives at a simple round-about after crossing a rhomb pattern
where the locomotive is doubled: at this point a second locomotive is appended to the arriving
one. At the first branching it meets, one locomotive is deleted. At the second one, the
single locomotive is sent on the right track. 

\vtop{
\vspace{-5pt}
\ligne{\hfill
\includegraphics[scale=0.5]{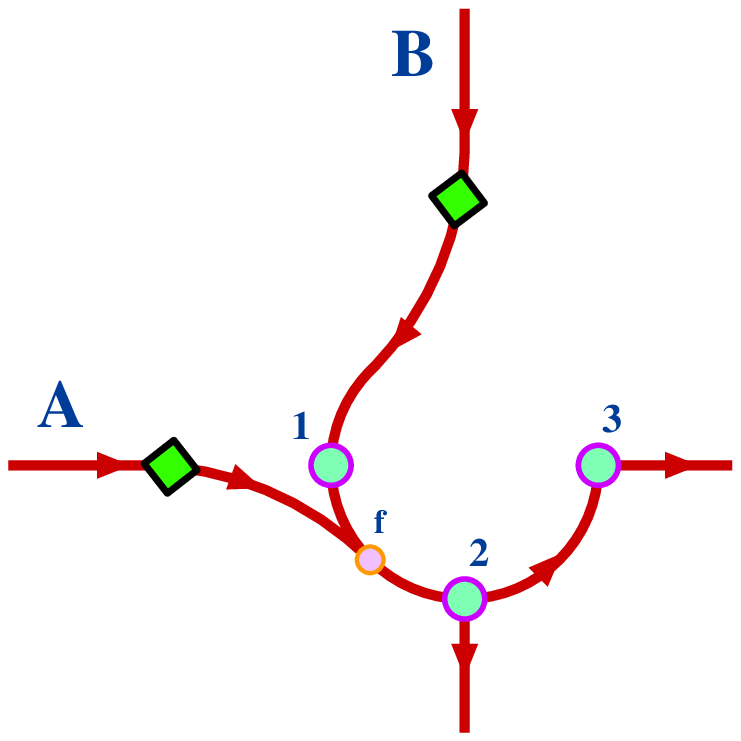}
\hfill}
\vspace{-20pt}
\begin{fig}\label{rondpointsimple}
\leurre
The new crossing: the one-way tracks from{\bf A} and {\bf B} intersect. We have a three-quarters 
round-about. 
The small disc at~{\bf\small f} represents a fixed switch. Discs~{\bf 1}, {\bf 2} and~{\bf 3}
represent the pattern which dispatches the motion of the locomotive on the appropriate way.
Patterns~{\bf 1} and~{\bf 3} are needed as explained in the description of the scenario.
\end{fig}
}

\vtop{
\vspace{-25pt}
\ligne{\hfill
\includegraphics[scale=0.5]{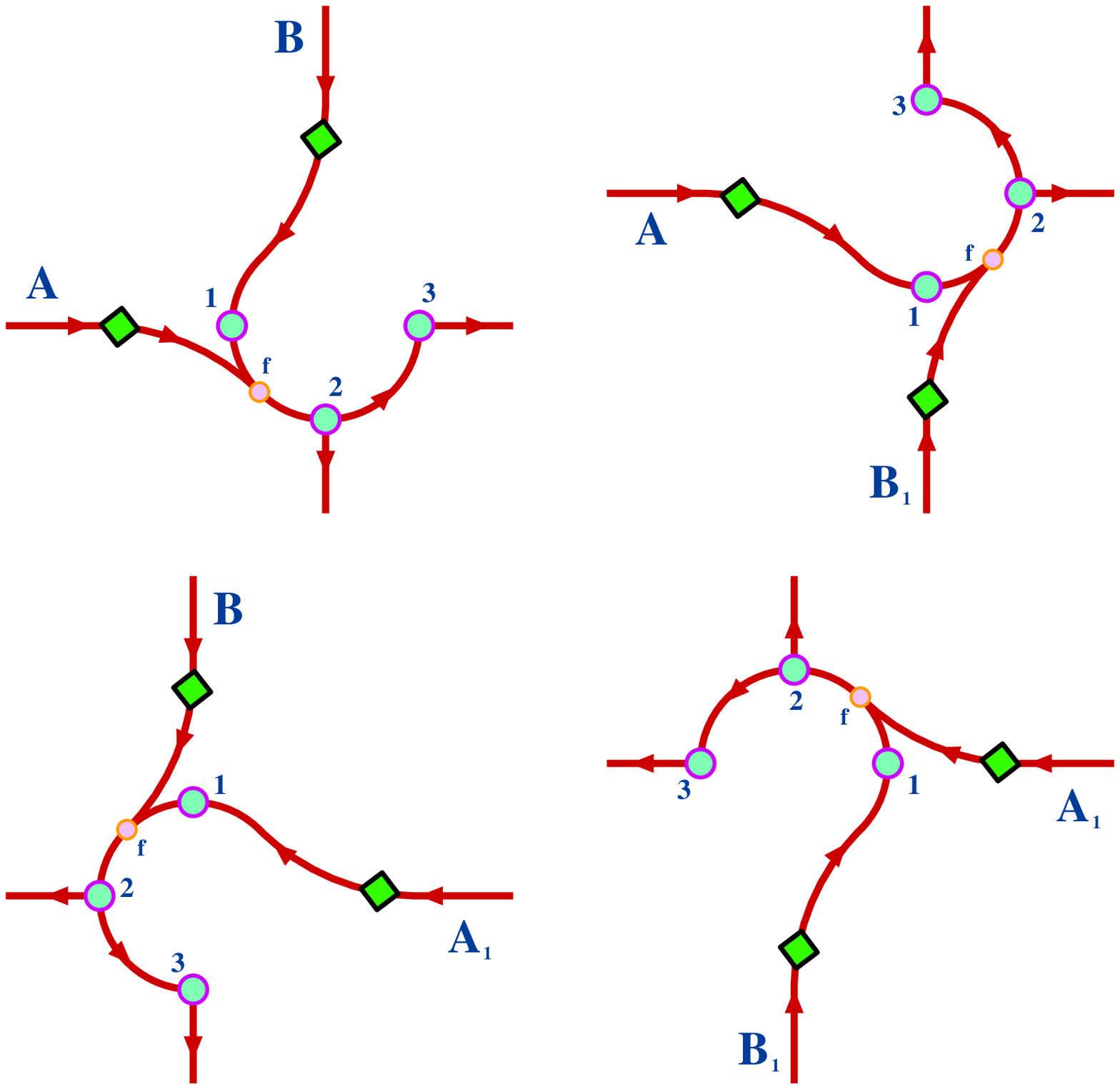}
\hfill}
\vspace{-10pt}
\begin{fig}\label{rondpointcomplet}
\leurre
The new crossing: four possible one-way track. Assembling them allows to perform a
two-way crossing. The notations are those of Figure~{\rm\ref{rondpointsimple}}. {\bf A}$_1$,
{\bf B}$_1$ go opposite to~{\bf A}, {\bf B}, respectively.
\end{fig}
\vspace{10pt}
}

   Figure~\ref{rondpointcomplet} shows us how to assemble four one-way simple round-abouts in 
order to perform a true crossing for two intersection two-ways tracks. The sturcture
of Figure~\ref{rondpointcomplet} is called a {\bf full round-about}.

But this change is not enough. We have to device a new implementation
for both the flip-flop and the memory switch to which we now turn.

   This new scenario for the implementation of these switches is a consequence of the
reduction of the number of states to~3 ones. The idea is to put in a common structure
the different working of the active part of the memory switch and the flip-flop.
In fact, we have to dissociate the dispatching of the locomotive from the control of the
selected track.

   For this purpose, we introduce two new structures: the {\bf fork} and the {\bf killer}.
The fork looks like a switch to which the locomotive arrives from the track~$a$. But, after
crossing the central cell, two locomotives leave the fork: one through the tack~$b$, the other
through the track~$c$. Now, the role of the killer is to kill one of these locomotives as
a single locomotive leaves a switch. We simply take advantage that the selection may happen a bit
further from the switch: this allows us to relax the crossing of the fork leaving two locomotives.
Now, the killer has two positions: an active one and a passive one. When the killer is active,
if a locomotive arrives at the killer, it is destroyed while it crosses its structure.
When the killer is passive, it let the locomotive leave its structure. We can see that it is enough
to construct the killer in such a way that a signal arriving to it changes its status from active
to passive and from passive to active.

\vtop{
\vspace{-15pt}
\ligne{\hfill
\includegraphics[scale=0.6]{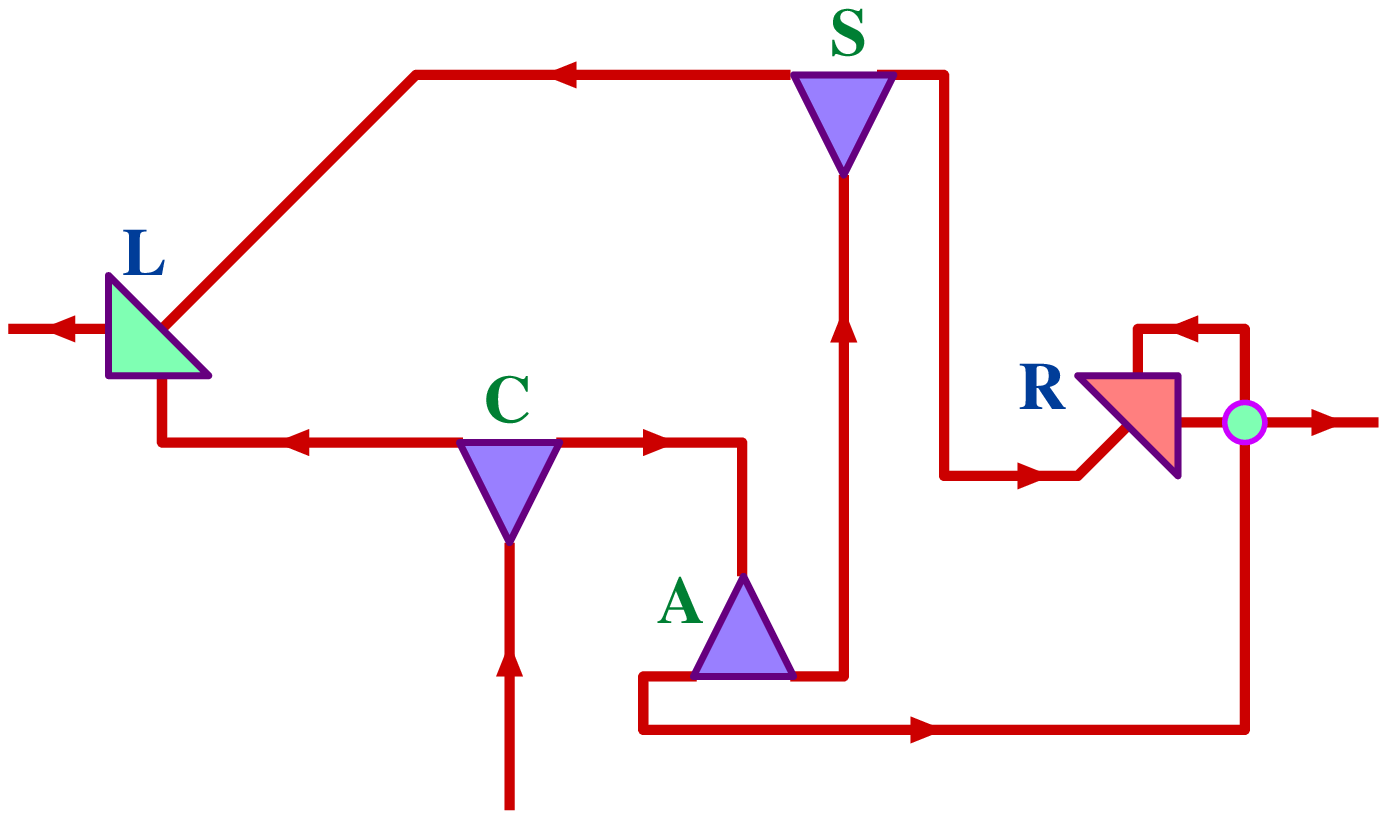}
\hfill}
\vspace{-10pt}
\begin{fig}\label{bascul_73}
\leurre
The flip-flop. Note the forks $C$, $A$ and~$S$ and the killers~$L$ and~$R$. Note the intersection
close to~$R$ which need a simple round about only.
\end{fig}
\vspace{10pt}
}

   Figure~\ref{bascul_73} shows us how we can assemble three forks and two killers in order
to get a flip-flop. Indeed, the locomotive arrives at~$C$ as indicated by the arrow below
the fork. Then, two locomotive leave the fork. The left-hand side locomotive~$\ell$ arrives to 
the killer~$L$ which, in the figure, is assumed to be passive. Accordingly, $\ell$~goes on its way
on the circuit. The right-hand side locomotive~$r$ arrives to a new fork, $A$. Then, two locomotives
leave~$A$. Call~$r$ the one which goes to the left and~$s$ the one which goes to the right.
We can see that $r$~arrives to the killer~$R$ which is in the other configuration than the 
killer~$L$. Accordingly, $R$~is active and it kills the locomotive~$r$. This means that
the locomotive is now~$\ell$. Let us look at what happens with~$s$. This locomotive is sent to
a new fork~$S$. From there, two locomotives leave the fork, $s_\ell$ to the left, $s_r$ to 
the right. We can see that $s_\ell$, $s_r$ arrives to~$L$, $R$ respectively. Accordingly, both
killers change their configuration~: $L$ becomes active and $R$~becomes passive. This
is exactly what a flip-flop has to do.

   Figure~\ref{memoact_73} represents the assembling of forks and killers in order to get an
active memory switch. Here, we can see that the locomotive arriving to~$C$ triggers the sending
of two locomotives~$\ell$ and~$r$, to left and to right respectively. We can see that $\ell$~crosses
a passive killer so that it goes on its way and will become the continuation of the locomotive
which arrived to~$C$. On the other hand, $r$~arrives to~$R$ which is active, so that~$r$ is 
cancelled at this moment. Now, the killers do not change their configuration. The configuration
is changed only when a locomotive arrives to~$S$, sent from the passive memory switch, in order
to change the selection of the switch.

\vtop{
\vspace{-15pt}
\ligne{\hfill
\includegraphics[scale=0.6]{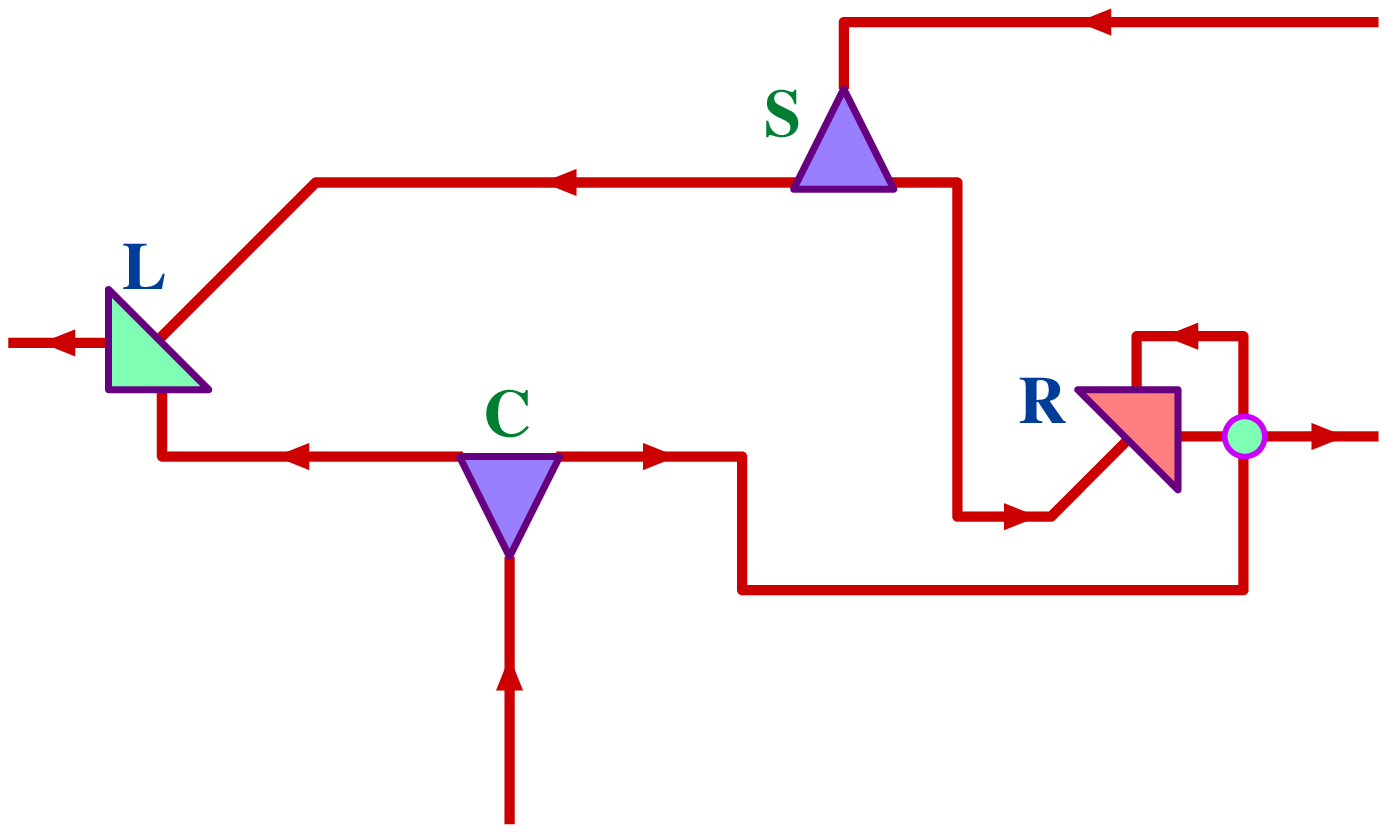}
\hfill}
\vspace{-10pt}
\begin{fig}\label{memoact_73}
\leurre
The active memory switch. Note the forks $C$ and~$S$ and the killers~$L$ and~$R$.
\end{fig}
\vspace{10pt}
}

\newdimen\largea\largea=16pt
\newdimen\largeab\largeab=32pt
\newdimen\largebb\largebb=42pt
\def\laligne #1 #2 #3 #4 #5 #6 {
\hbox to \largeab{\hfill#1\hfill}
\hbox to \largea{\hfill#2\hfill}
\hbox to \largea{\hfill#3\hfill}
\hbox to \largea{\hfill#4\hfill}
\hbox to \largea{\hfill#5\hfill}
\hbox to \largea{\hfill#6\hfill}
}
\def\lautreligne #1 #2 #3 #4 #5 {
\hbox to \largebb{\hskip 10pt\hfill#1\hfill}
\hbox to \largea{\hfill#2\hfill}
\hbox to \largea{\hfill#3\hfill}
\hbox to \largea{\hfill#4\hfill}
\hbox to \largea{\hfill#5\hfill}
}
\vtop{
\begin{tab}\label{fork_kill}
Scheme of execution of the flip-flop and the active memory switch.
\end{tab}
\ligne{\hfill
\laligne {} {$C$} {$A$} {$S$} {$L$} {$R$} \lautreligne {} {$C$} {$S$} {$L$} {$R$} \hfill}
\ligne{\hfill
\laligne {0} {$-$} {$-$} {$-$} {$b$} {$r$} \lautreligne {0} {$-$} {$-$} {$b$} {$r$} 
\hfill}
\ligne{\hfill
\laligne {1} {$\bullet$} {$-$} {$-$} {$b$} {$r$} \lautreligne {1} {$\bullet$} {$-$} {$b$} {$r$} 
\hfill}
\ligne{\hfill
\laligne { } {$-$} {$-$} {$-$} {$b$} {$r$} \lautreligne {} {$-$} {$-$} {$b$} {$r$} 
\hfill}
\ligne{\hfill
\laligne {2} {$-$} {$\bullet$} {$-$} {$\bullet b$} {$r$} \lautreligne {2} {$-$} {$\bullet$} 
{$\bullet b$} {$\bullet r$} 
\hfill}
\ligne{\hfill
\laligne {} {$-$} {$-$} {$-$} {$\bullet b$} {$r$} \lautreligne {} {$-$} {$-$} 
{$\bullet b$} {$- r$} 
\hfill}
\ligne{\hfill
\laligne {3} {$-$} {$-$} {$\bullet$} {$b$} {$\bullet r$} \lautreligne {3} {$-$} {$-$} {$b$} 
{$r$} \hfill}
\ligne{\hfill
\laligne {} {$-$} {$-$} {$-$} {$b$} {$- r$} \lautreligne {} {$-$} {$-$} {$b$} {$r$} \hfill}
\ligne{\hfill
\laligne {4} {$-$} {$-$} {$-$} {$\nabla b$} {$\nabla r$} \lautreligne {$n$} {$-$} {$\bullet$} 
{$b$} {$r$} \hfill} 
\ligne{\hfill
\laligne {} {$-$} {$-$} {$-$} {$r$} {$b$} \lautreligne {} {$-$} {$-$} 
{$b$} {$r$} \hfill} 
\ligne{\hfill
\laligne {} {} {} {} {} {} \lautreligne {$n$+1} {$-$} {$-$} 
{$\nabla b$} {$\nabla r$} \hfill} 
\ligne{\hfill
\laligne {} {} {} {} {} {} \lautreligne {} {$-$} {$-$} 
{$r$} {$b$} \hfill} 
\vskip 10pt
}

   Table~\ref{fork_kill} summarizes the working we above indicated. Note that in each sub-table,
the leftmost column indicates a time. It should be stressed that between two such times, the 
number of steps of the cellular automaton may be very different. The only requirement is that
before the visit of the next site by the locomotive, switch or crossing, all changes involved 
by the current site have been completed. In particular, when $S$~triggers two locomotive in order to
change the status of both~$L$ and~$R$, there is no need that the change happens at the same
step of the cellular automaton. It is enough that when the locomotive arrives to the next site,
the change has been performed. Note that we also can consider that the changes at a site are 
completed when there is again exactly one locomotive on the circuit. While performing the crossing
of a flip-flop, there can be up to three locomotives running on the circuit at the same instant.
When the flip-flop has changed the selection of the tracks, there remains exactly one locomotive.

   Now that we have seen the scenario to implement the circuit, we have to precisely look
at how to implement it with three states only. We turn now to this question.

\subsubsection{Implementing the new scenario}
\label{scenar}
\def\YY{\hbox{\tt Y}{}}
   As already announced, we need to use three states only. The first state is the quiescent state
which we denote by \WW. Remember it is defined by the following rule: if a cell~$c$ and all
its neighbours are in the state~\WW, then at the next top of the clock, the cell~$c$ remains
in the state~\WW. We shall also call~\WW{} the blank and we also shall say that a cell in~\WW{}
is white. The other states are~\BB{} and  \RR. The cells in these states are said
to be blue and red, respectively. The state~\BB{} is mainly used for the 
milestones which delimit the tracks. The state~\RR{} is the basic state of the locomotive,
remember that now it only consists of one cell. In several implementations, the marking of
the structure also makes use of the state~\RR.
 
  Checking the new scenario requires to first study the implementation of the tracks. We have to
define verticals and horizontals.

\vskip 7pt
\noindent
\ligne{$\underline{\hbox{Vertical and horizontal tracks}}$}
\vskip 5pt

  The motion is basically constructed by motions around a cell, the {\bf pivot}, with two kinds 
of links from one pivot to another in a way which is slightly different from what we did in
the grid $\{13,3\}$. Figure~\ref{elemvoie} shows the elements of the track and, for each of them,
how it is crossed by the locomotive.

 To better understand the figure, number the sides of the central cell as follows.
Number the sides increasingly while counter-clockwise turning around the cell. The milestones
occur on the sides~1, 3 and~6{} in the first row, 1~being the side shared by the leftmost
milestone.

   The first element is the {\bf standard} one, on the first row of the figure. 
The second element is the {\bf turn},
on the second row of the figure. 
There the
locomotive goes from side~7 to side~2, exactly the reverse motion of the previous one
and the milestones are on sides~1, 3, 4 and~6. The last 
element is the {\bf junction}, on the third row of Figure~\ref{elemvoie}, 
There, the locomotive goes from side~5 to side~2, with milestones on sides~1, 3, 4,
6 and~7. In the first row of 
Figure~\ref{elemvoie}, we can say that the pivot is the blue cell sharing side~1. As will be seen
in Section~\ref{rules}, we can append rules so that the same motion is possible for each
of these elements with two contiguous red cells instead of a single one.

In Figure~\ref{voievert} we make use of the standard element and of the turn in order to
perform tracks travelling along a branch of a Fibonacci tree: we call this track a {\bf vertical}.
As can be seen on the figure, the standard elements are turning around a pivot, running along a 
half of its sides and passing from one pivot to the other thanks to the turn. The turn is 
absolutely needed in order to obtain tracks which can go from any tile to any other one.

\vtop{
\vspace{-15pt}
\ligne{\hfill
\includegraphics[scale=0.3]{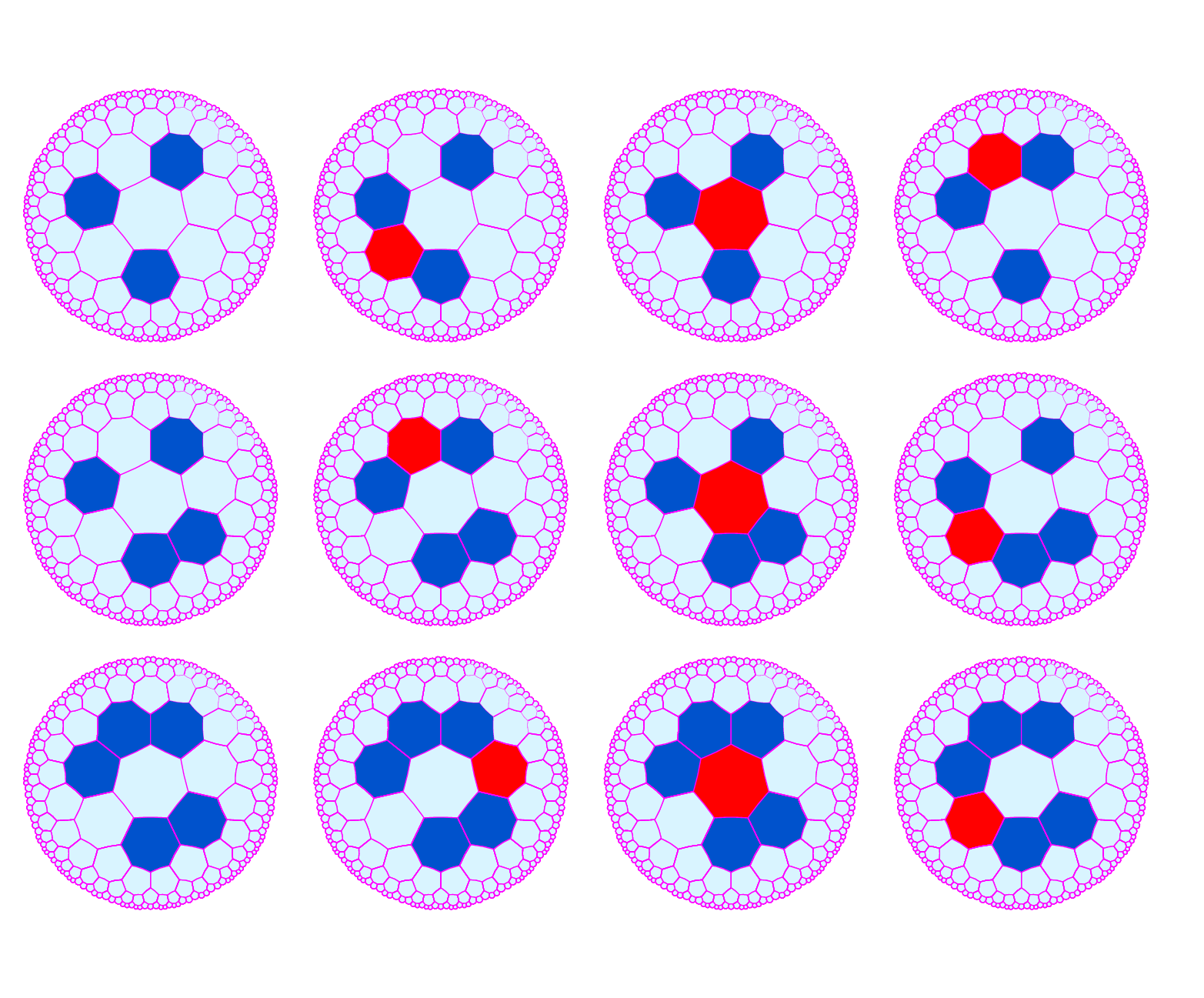}
\hfill}
\vspace{-20pt}
\begin{fig}\label{elemvoie}
\leurre
The elements of the track. For each of them, the figures show how it is crossed by the
locomotive. From top to bottom: the standard element, the turn and the junction.
\end{fig}
\vspace{10pt}
}

\vtop{
\vspace{-5pt}
\ligne{\hfill
\includegraphics[scale=0.27]{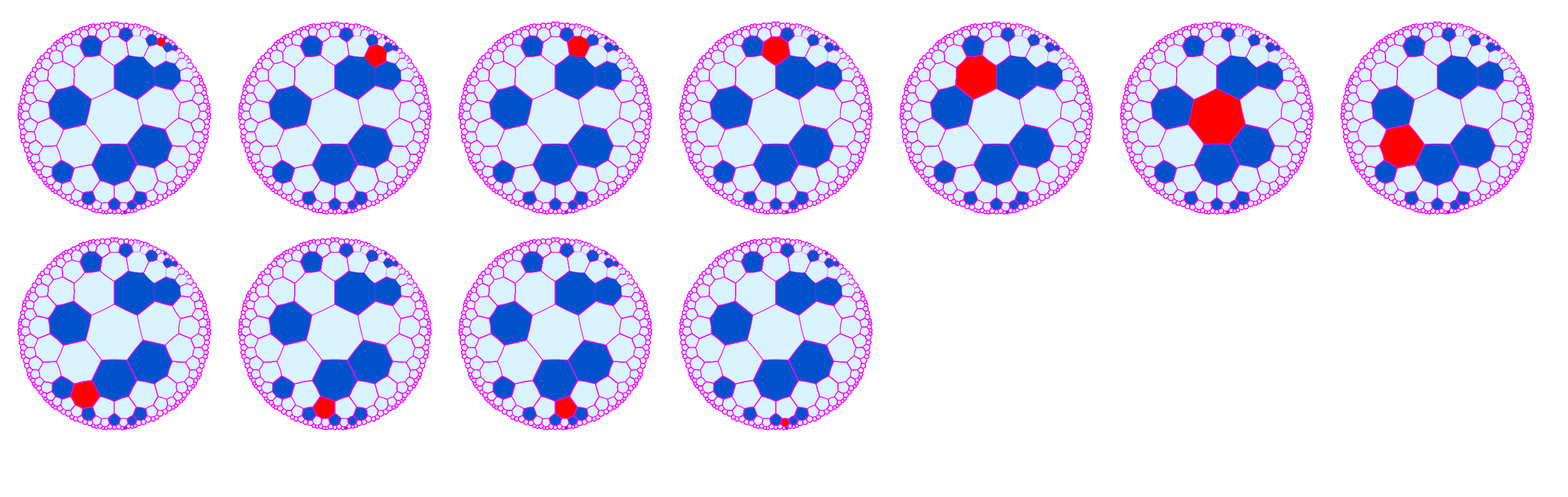}
\hfill}
\vspace{-10pt}
\ligne{\hfill
\includegraphics[scale=0.27]{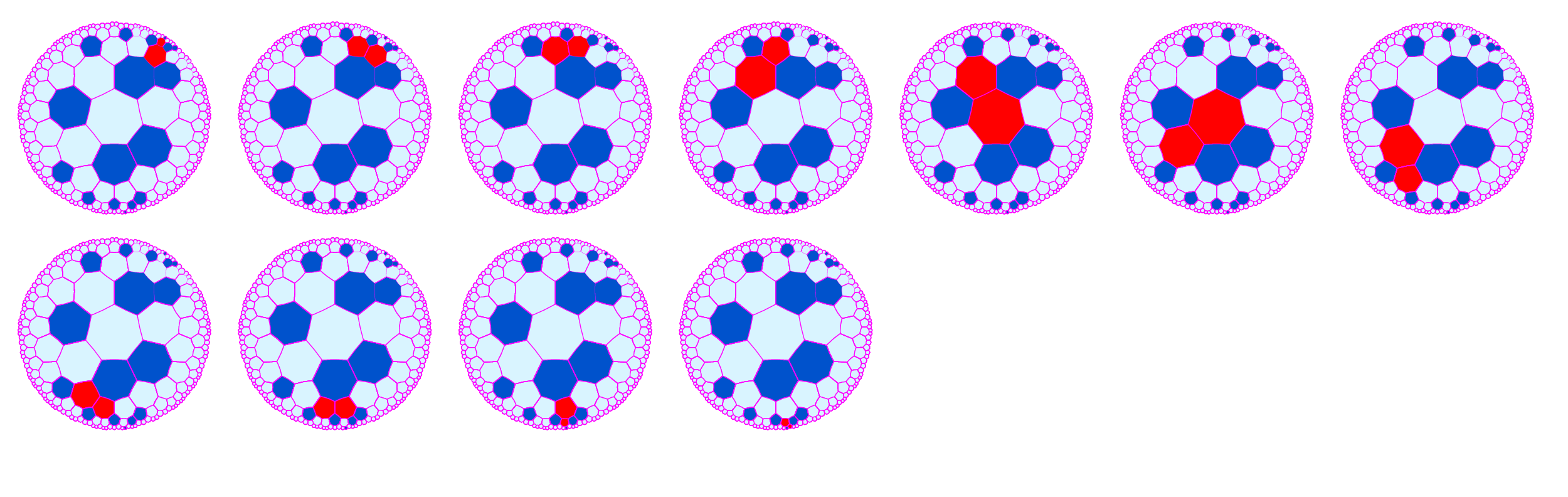}
\hfill}
\vspace{-20pt}
\begin{fig}\label{voievert}
\leurre
A vertical motion for the locomotive.
First two rows: a single locomotive; last two rows: two contiguous locomotives travelling together.
\end{fig}
\vspace{10pt}
}

   In Figure~\ref{voiehoriz} we can see the implementation of what we call a {\bf horizontal}:
it is a track running along a level of a Fibonacci tree. The pivots are on such a level and the
track itself, where the locomotive runs, consists of the sons of the nodes which are on the level
of the pivots. To go from the sons of a node to the sons of the next node, we precisely need
the junction. As can be seen on the figure, the same motion can be managed with two contiguous
locomotives travelling on the same path.

   As seven Fibonacci trees whose root are displayed around a fixed tile constitute a circular
path of any radius, combining this with verticals as also show in Figure~\ref{voiehoriz}, we
can see that we indeed can construct a path from a tile to another: we construct a shortest path
whose tiles play the role of pivots and we go from one pivot to the next one through
a turn or through a junction, depending on the configuration. 

\vtop{
\vspace{-10pt}
\ligne{\hfill
\includegraphics[scale=0.27]{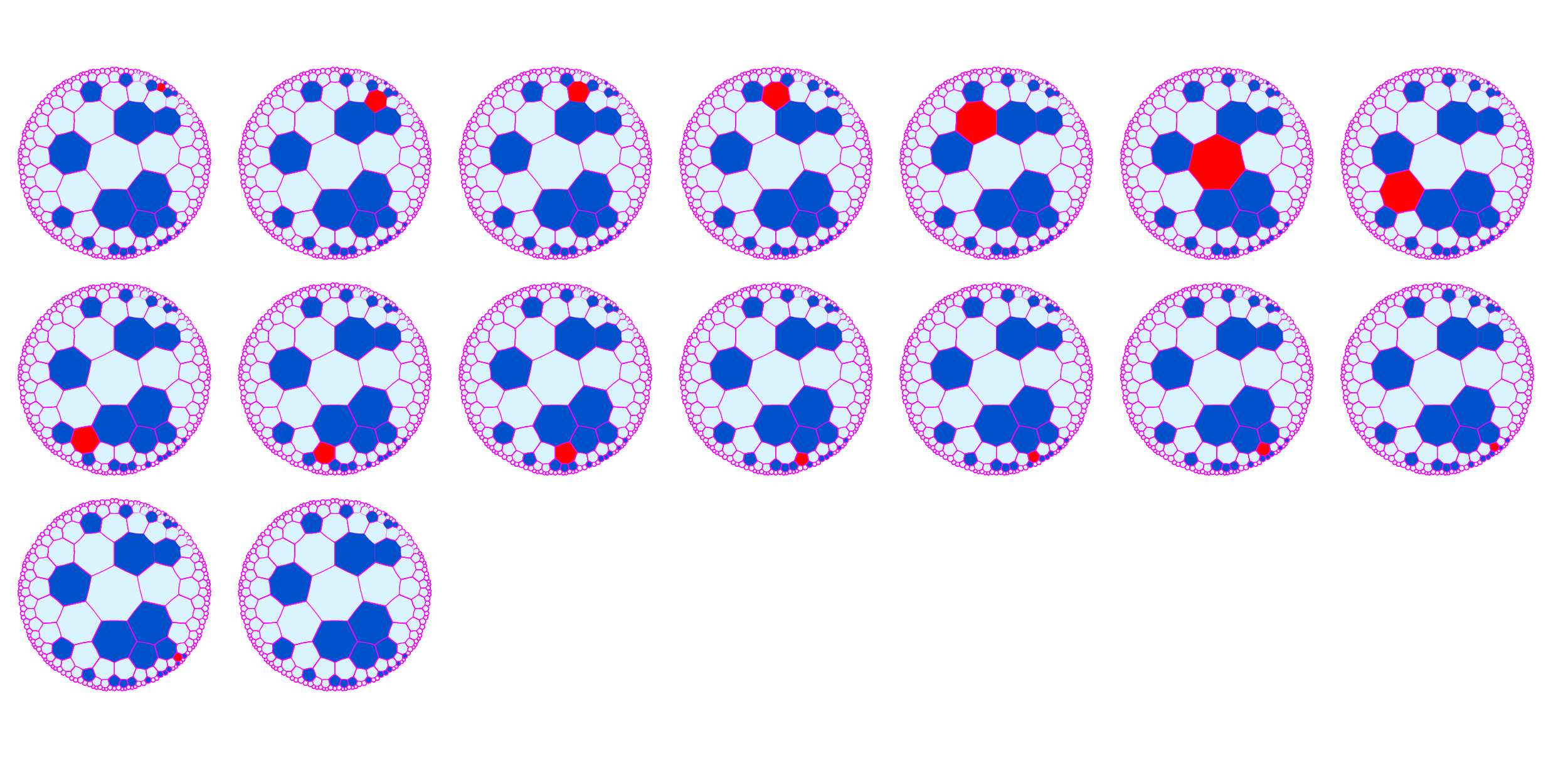}
\hfill}
\vspace{-20pt}
\ligne{\hfill
\includegraphics[scale=0.27]{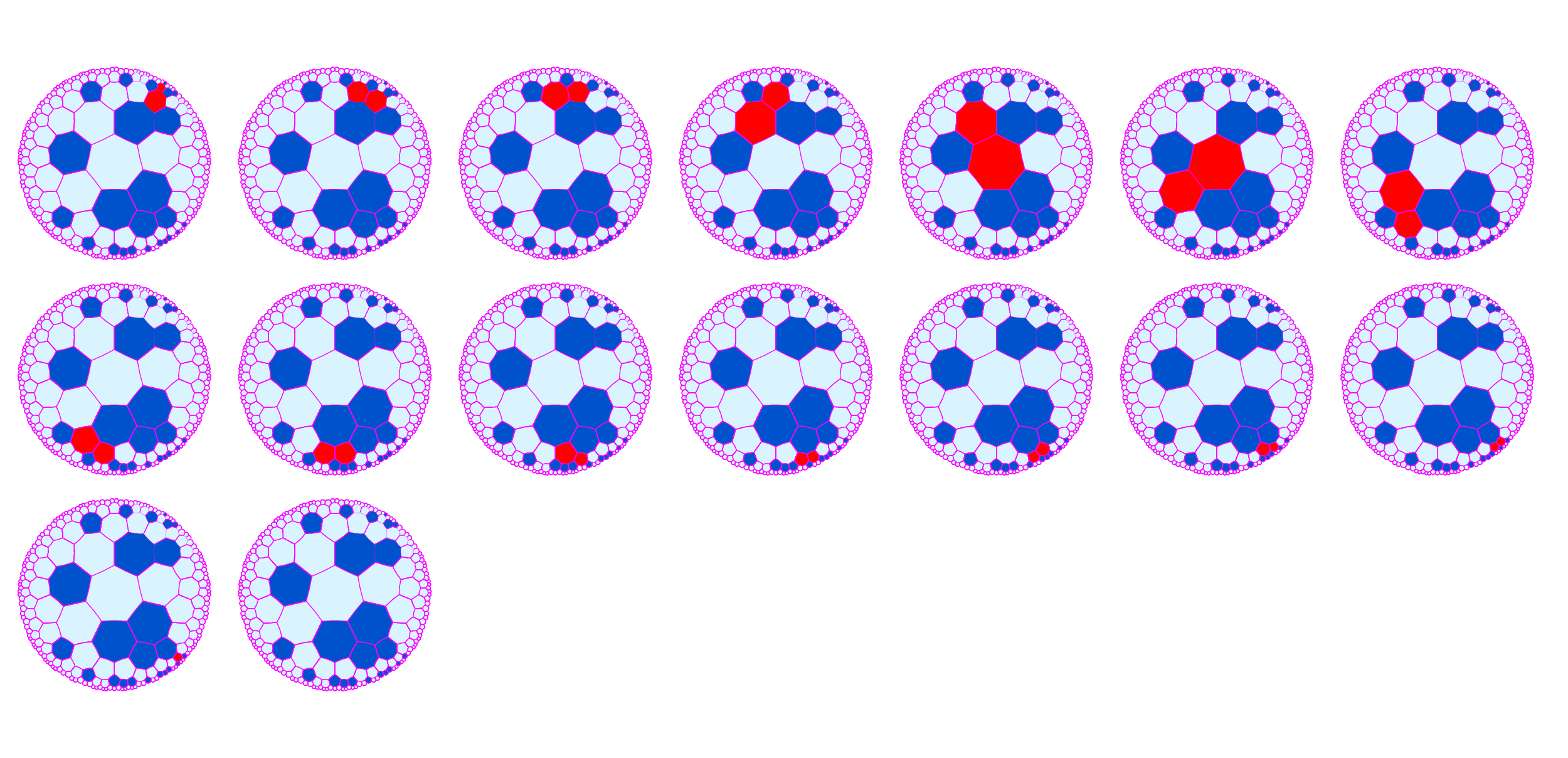}
\hfill}
\vspace{-15pt}
\begin{fig}\label{voiehoriz}
\leurre
A horizontal with a locomotive, first four rows and then, two contiguous locomotives.
\end{fig}
\vspace{10pt}
}

   Remark that as horizontal tracks run on three consecutive levels, a two-way
portion of the tracks require at least seven consecutive levels as we need at least one level to
separate the tracks run on each direction. Similarly,
a two-way section of vertical lines must be separated by several nodes on the same level at 
the level where the
distance between the supporting line is minimal: this guarantees 
that the lines are not secant. These constraints require much space for the implementation, but
in the hyperbolic plane, we are never short of space.

\vskip 7pt
\noindent
\ligne{$\underline{\hbox{Implementing the crossings}}$}
\vskip 5pt

   As indicated in Subsubsection~\ref{scenar}, the tracks are organized according to
what is depicted in Figures~\ref{rondpointsimple} and~\ref{rondpointcomplet}. From the
latter figure, it is enough to focus on the implementation of Figure~\ref{rondpointsimple}.
From the implementation of the tracks, we only have to look at the implementatation
of the patterns symbolically denoted as~{\bf 1}, {\bf 2}, {\bf 3}, {\bf f} and by a rhomb
in the figure. As {\bf f} is a fixed switch whose implementation is indicated a bit further,
we simply implement~{\bf 1} and the rhomb as {\bf 2} and {\bf 3} are strict copies of {\bf 1}.
Figure~\ref{hca73croisav} illustrates this implementation and the behaviour of the locomotive
when it crosses pattern~{\bf 1}. We can see that the difference strongly depends on whether
a single locomotive or two of them arrive to the pattern.

   From Figure~\ref{rondpointsimple}, the locomotive always arrives to the pattern from the same
cell, namely 9(5). Then the locomotive goes to the cell~1(6) if it is alone. If two contiguous
locomotives arrive, then a single one arrives to the cell~0 and then goes on its way on the
round-about. First, Figure~\ref{hca73croisdb} illustrates the working of the rhombic pattern
which duplicates a locomotive arriving to the pattern.

\vtop{
\vspace{-5pt}
\ligne{\hfill
\includegraphics[scale=0.27]{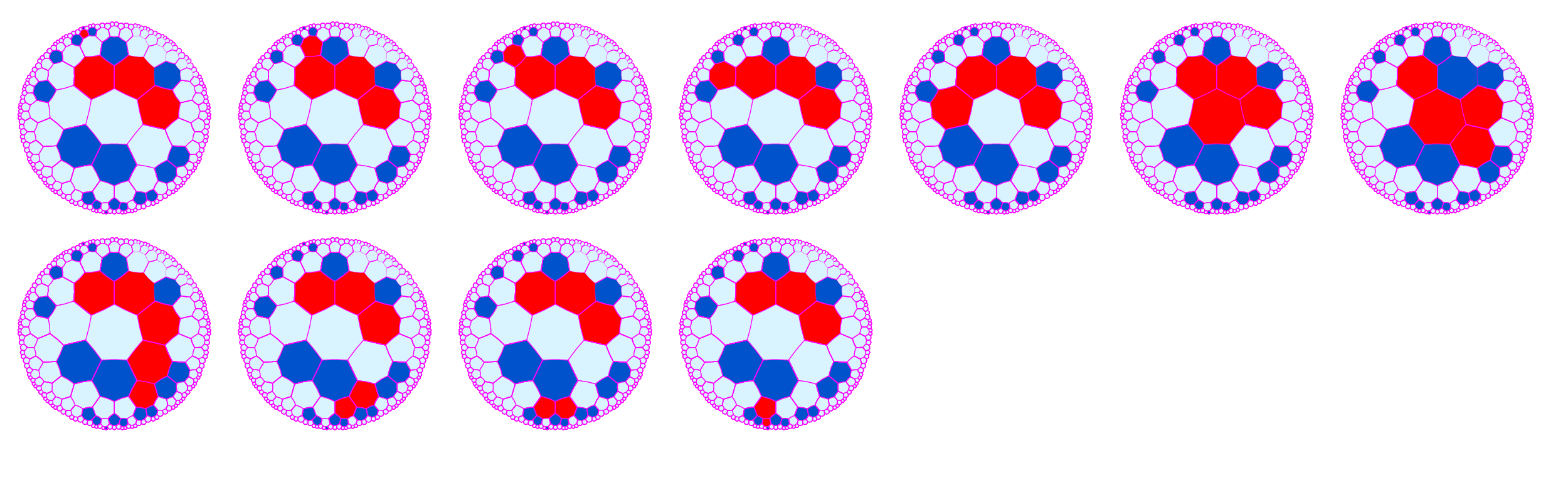}
\hfill}
\vspace{-15pt}
\begin{fig}\label{hca73croisdb}
\leurre
The rhombic pattern of the crossings. A single locomotive arrives at the pattern, two contiguous ones
leave the pattern.
\end{fig}
\vspace{10pt}
}

\vtop{
\vspace{-5pt}
\ligne{\hfill
\includegraphics[scale=0.27]{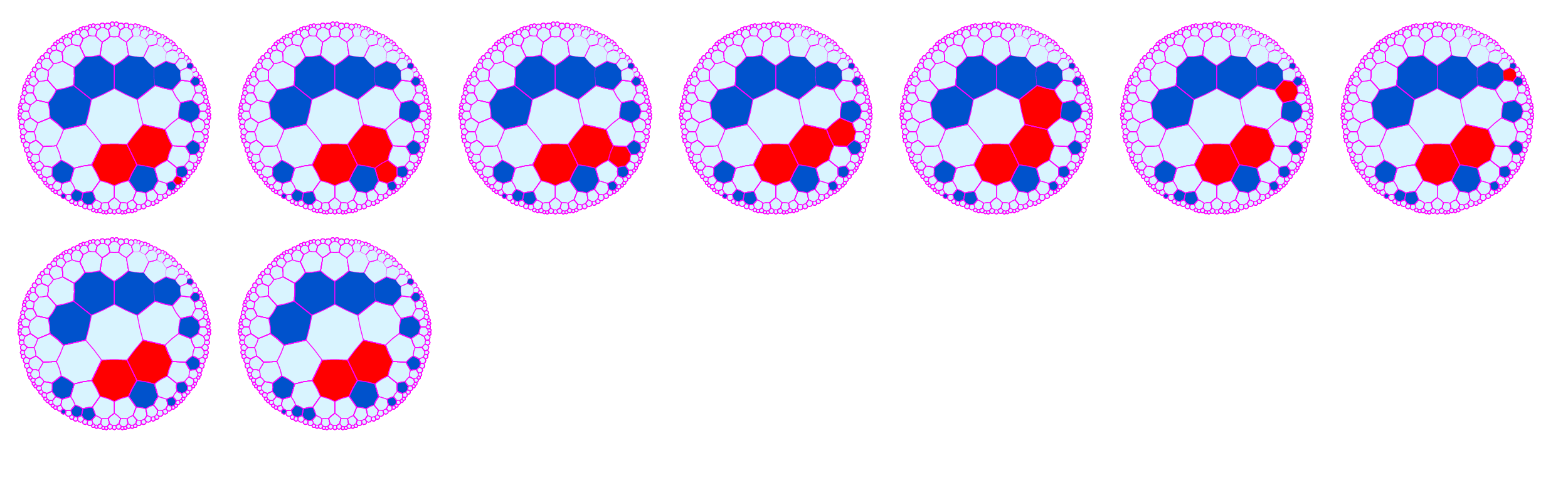}
\hfill}
\ligne{\hfill
\includegraphics[scale=0.27]{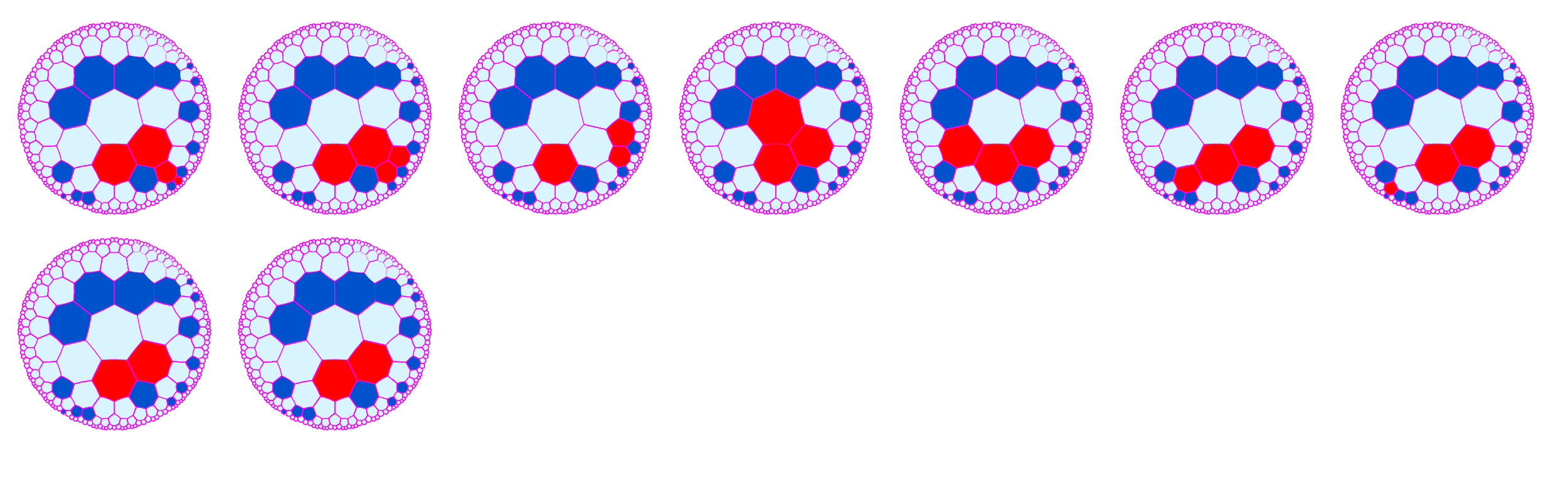}
\hfill}
\vspace{-15pt}
\begin{fig}\label{hca73croisav}
\leurre
The pattern {\bf 1} of the crossings. Notice the difference of behaviour depending on the
number of locomotives. When a single locomotive is detected, the locomotives goes to
the track. When two contiguous locomotives are detected, one of them is cancelled and the
other goes on its way on the round-about.
\end{fig}
\vspace{10pt}
}

   Next, Figure~\ref{hca73croisav} illustrates the pattern which is common to~{\bf 1}, {\bf 2}
and {\bf 3} in Figure~\ref{rondpointsimple}. We can see the difference of behaviour depending on
the number of contiguous locomotives which arrive at the pattern. If a single locomotive arrives
at the pattern, then the locomotive is sent to the track which leaves the round-about at this 
branching point. This is what is illustrated by the frist two rows of Figure~\ref{hca73croisav}.
If two contiguous locomotives arrive at the patter, then one locomotive is destroyed before it
is seen by the first cell of the track which leaves the round-about in the pattern. Then, the second
locmotive is sent to the way which continues the round-about. This is illustrated by the last
two rows of Figure~\ref{hca73croisav}.

\vskip 7pt
\noindent
\ligne{$\underline{\hbox{Implementation of a fixed switch}}$}
\vskip 5pt

   Figure~\ref{hca73fix} illustrates the working of the passive crossing of 
a fixed switch.

   The upper half of Figure~\ref{hca73fix} illustrates the case of a single locomotive.
In the first two rows of the figure, the locomotives arrives to the switch from the left. In the 
last two rows, it arrives from the right.

   Now, the lower half of Figure~\ref{hca73fix} illustrates the case of two contiguous 
locomotives. Again,
the first two rows of the figure illustrate the case when the locomotives arrive to the switch
from the left. The last two rows illustrate the case when these locomotives arrive to
the right.
   
With one-way tracks, we need a single kind of passive fixed switch. Indeed, there is no need
of an active fixed switch, as the locomotive never goes in the non-selected direction. The active
switch is reduced to the track which goes in the selected direction, as illustrated by
Figure~\ref{newswitches}.

   We remain with the more complicated switches: the flip-flop and the memory switch. 
Remember from Sub-section~\ref{newscenar} that we split the memory switch into two sub-switches,
the active memory switch and the passive one, due to the introduction of one way tracks for the
locomotive. In the same sub-section, we have seen that the flip-flop and the active memory switch
can be implemented by using the same structures, the fork and the killer: we only have to dispatch 
them in different way in order to implement the two kinds of switches. Now, we could probably
use the fork to implement the passive memory switch, but it would require another structure,
let us call it the {\bf sensor}. As we succeeded to implement a sensor with what is indeed another
fork in a single structure, We start with the passive memory switch.

\vtop{
\vspace{-20pt}
\ligne{\hfill
\includegraphics[scale=0.27]{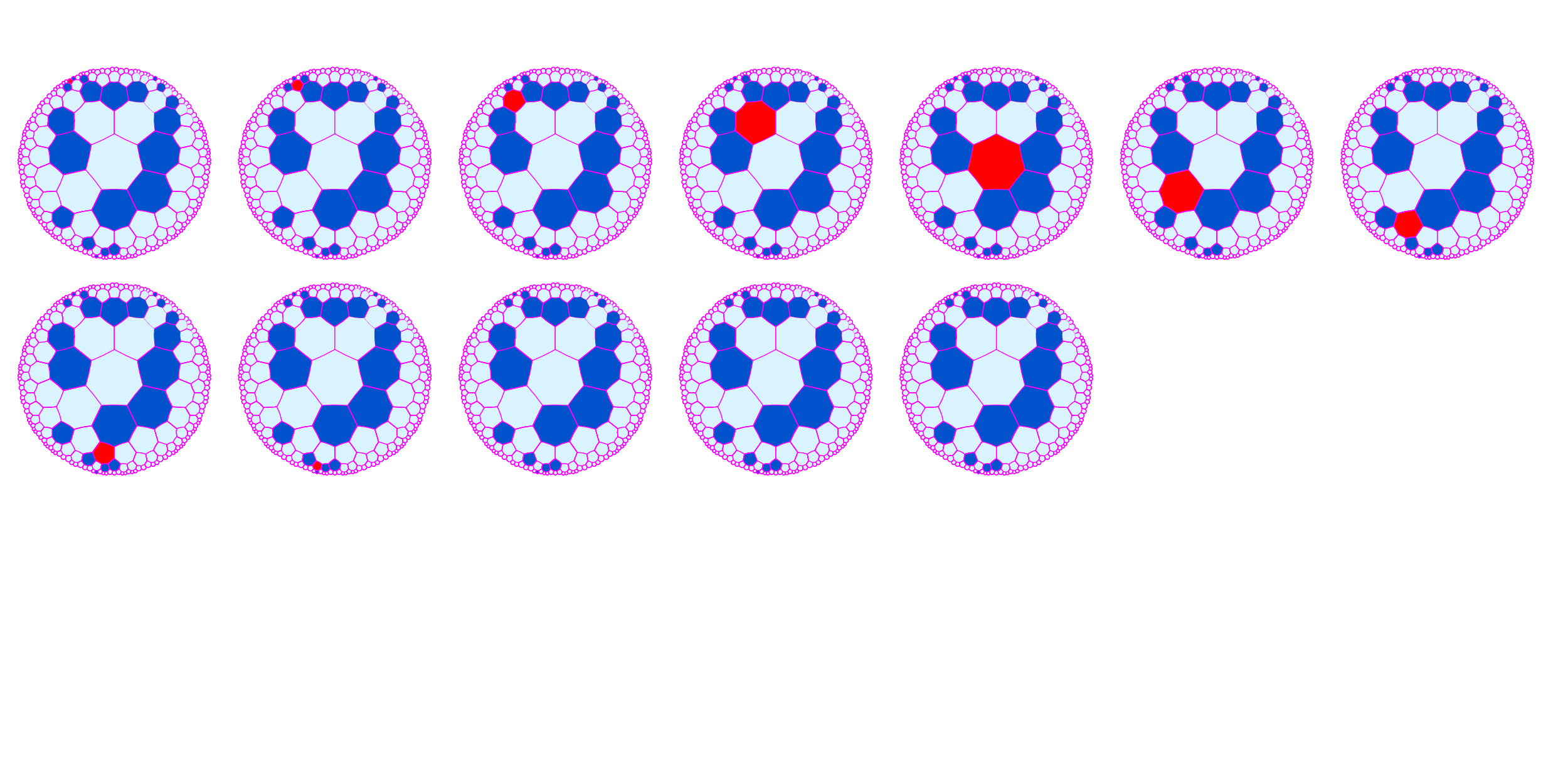}
\hfill}
\vspace{-110pt}
\ligne{\hfill
\includegraphics[scale=0.27]{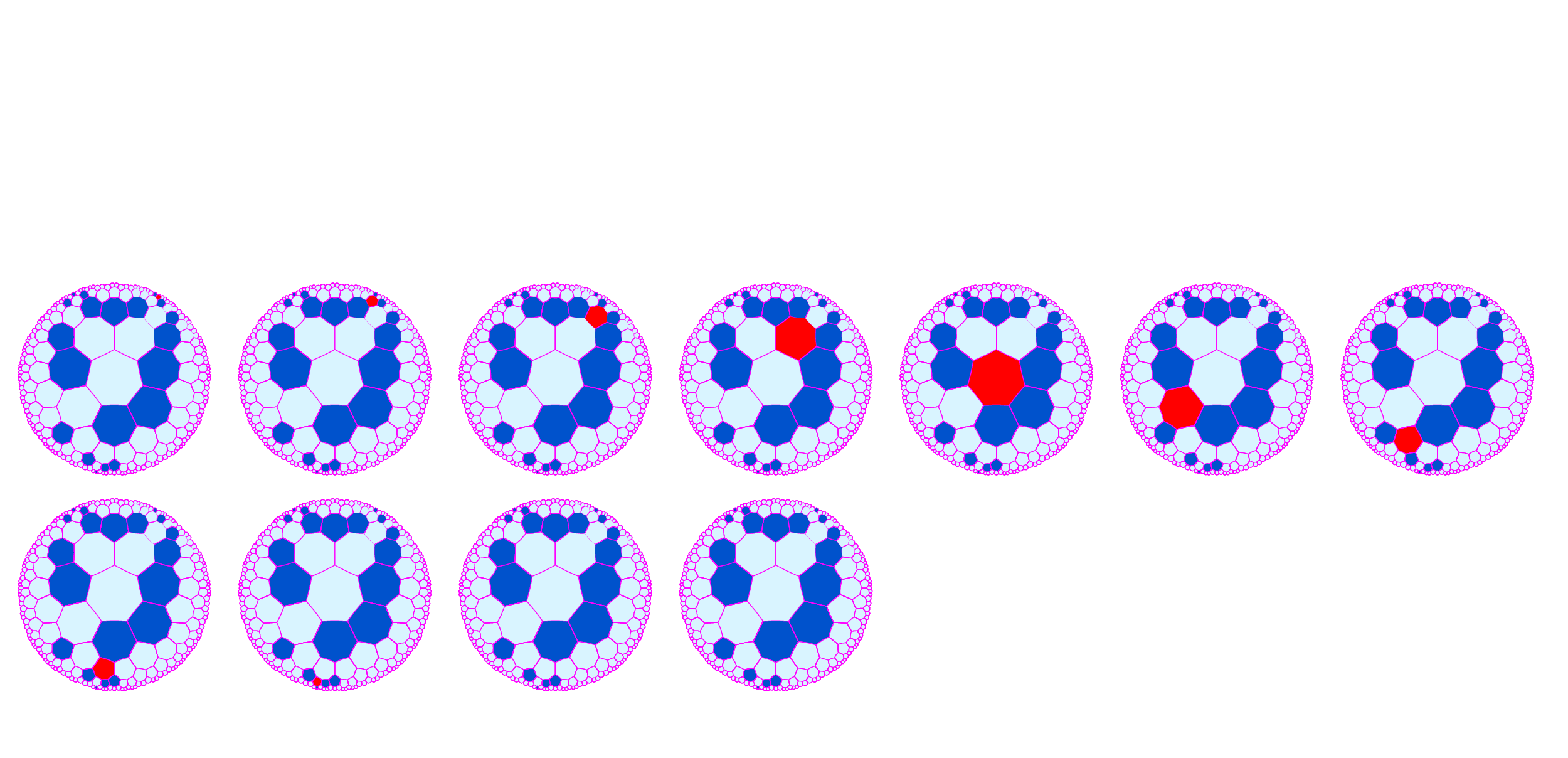}
\hfill}
\vspace{-20pt}
%
\ligne{\hfill
\includegraphics[scale=0.27]{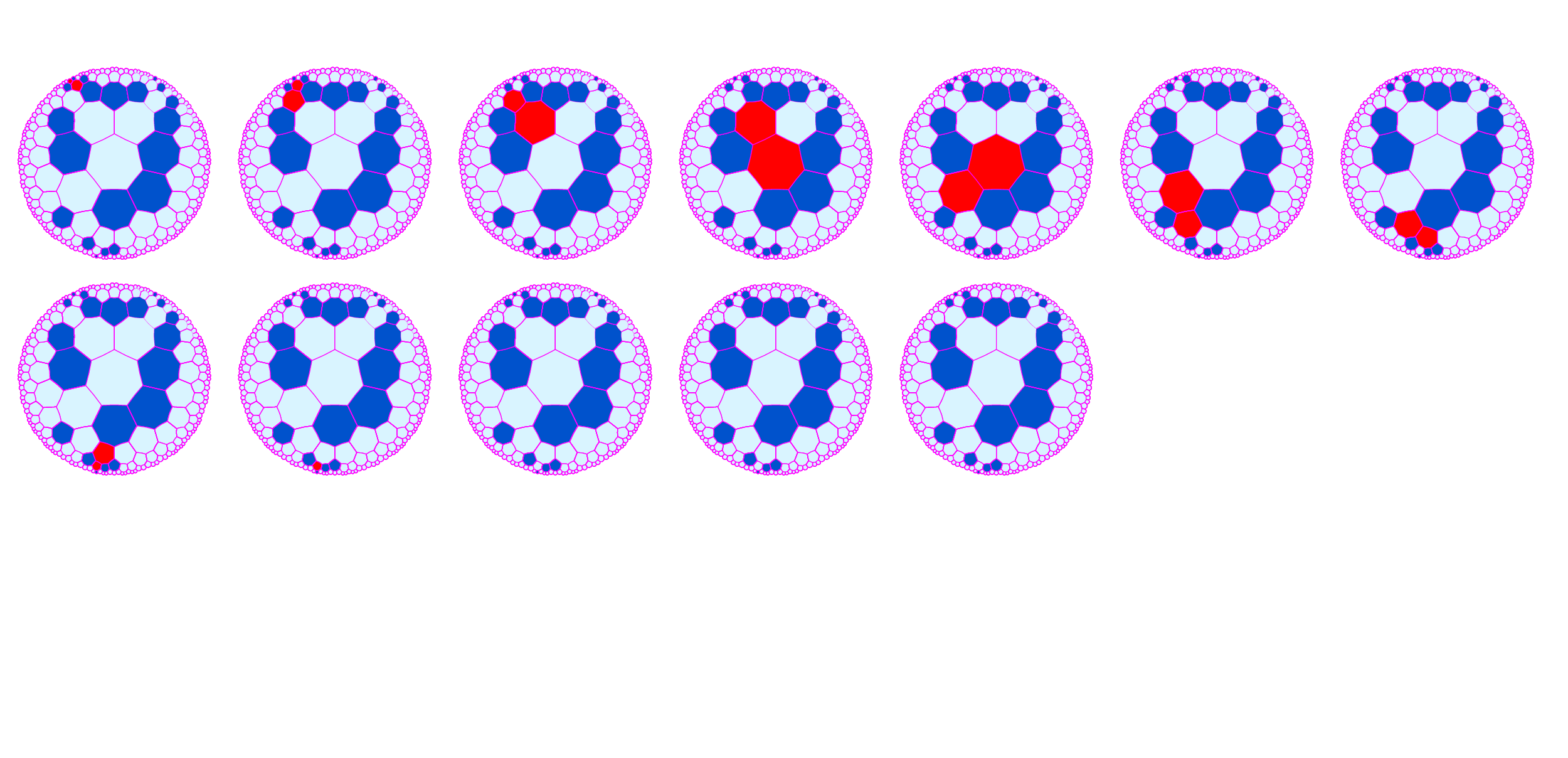}
\hfill}
\vspace{-110pt}
\ligne{\hfill
\includegraphics[scale=0.27]{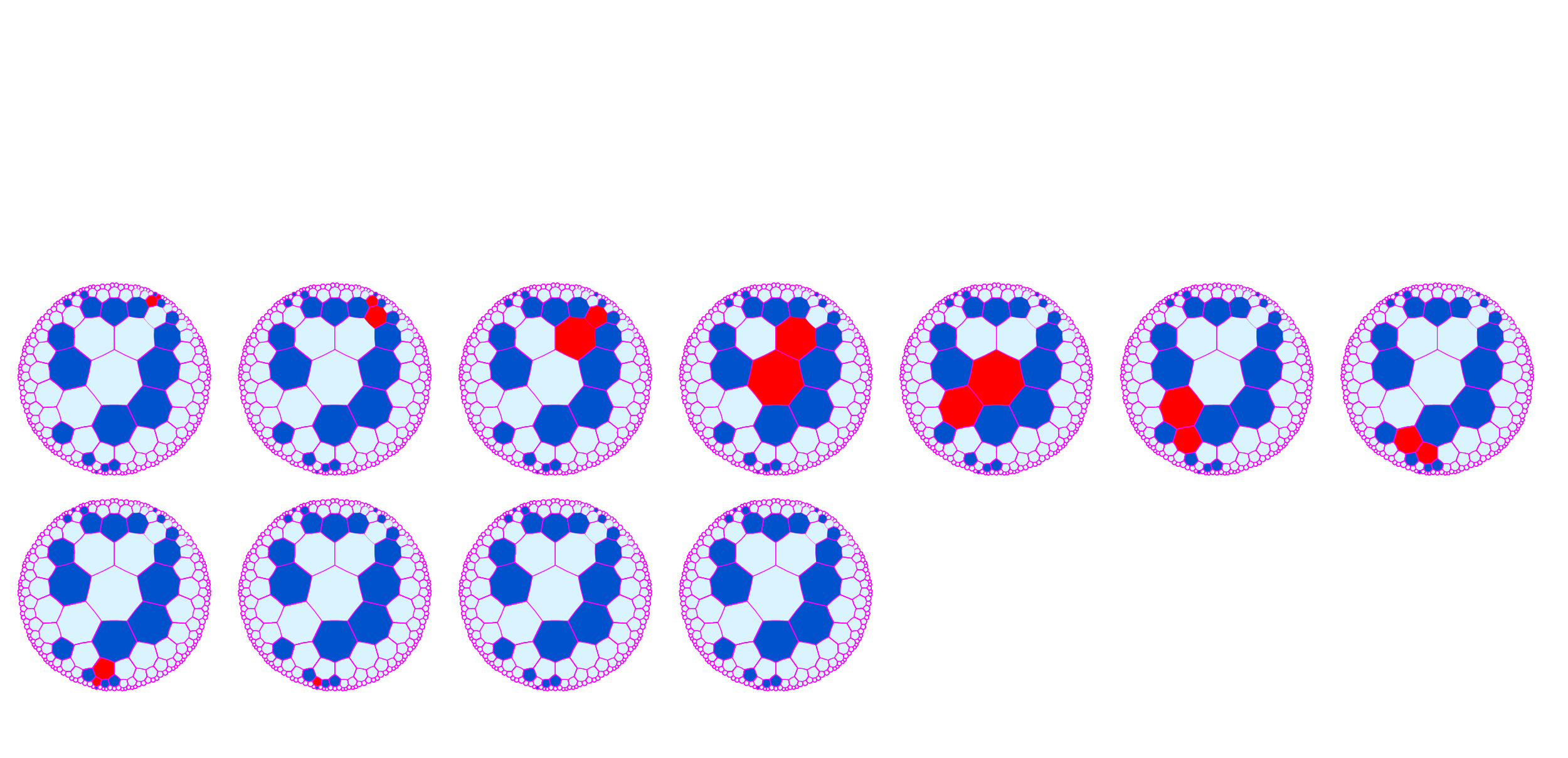}
\hfill}
\vspace{-20pt}
\begin{fig}\label{hca73fix}
\leurre
The fixed switch: passive crossing by the green locomotive and then the red one.
\end{fig}
\vspace{10pt}
}

\vskip 7pt
\noindent
\ligne{$\underline{\hbox{Implementation of a passive memory switch}}$}
\vskip 5pt

   Both idle configurations are illustrated by Figure~\ref{hca73mempassstables}: the left- and
the right-hand side switch.

\vtop{
\vspace{-15pt}
\ligne{\hfill
\includegraphics[scale=0.17]{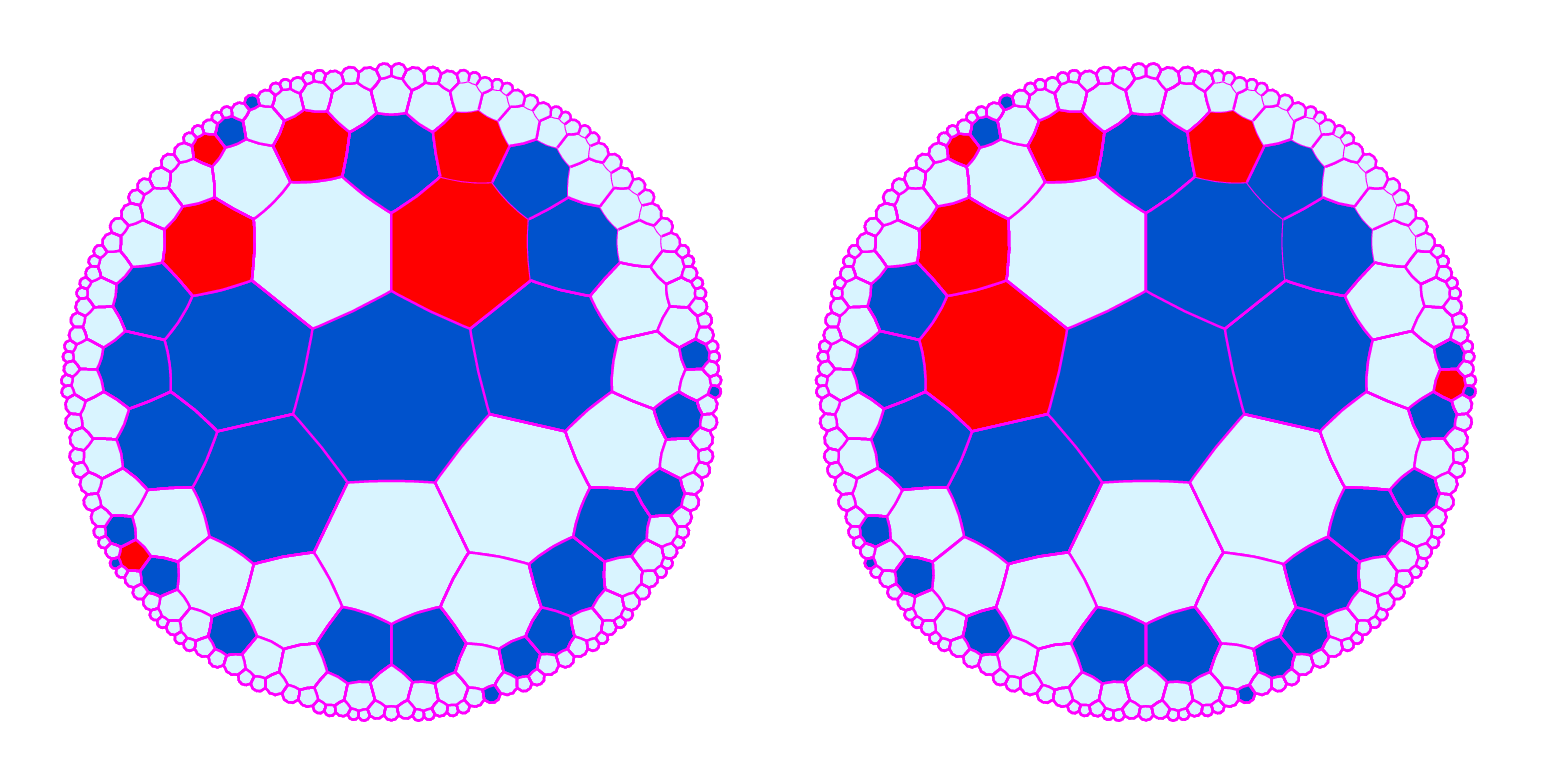}
\hfill}
\vspace{-10pt}
\begin{fig}\label{hca73mempassstables}
\leurre
The stable configurations of the passive memory switch: the side of the selected track is that
of the picture.
\end{fig}
}

   The working of this switch is illustrated by Figure\ref{hca73memopassif}. The upper half of the
figure illustrates the working of the left-hand side switch. In the first two lines, the locomotive
arrives from the left so that nothing happens. In the next two lines, the locomotive arrives from
the right-hand side~: this time the selection is changed to the right-hand side one.

\vtop{
\vspace{-45pt}
\ligne{\hfill
\includegraphics[scale=0.27]{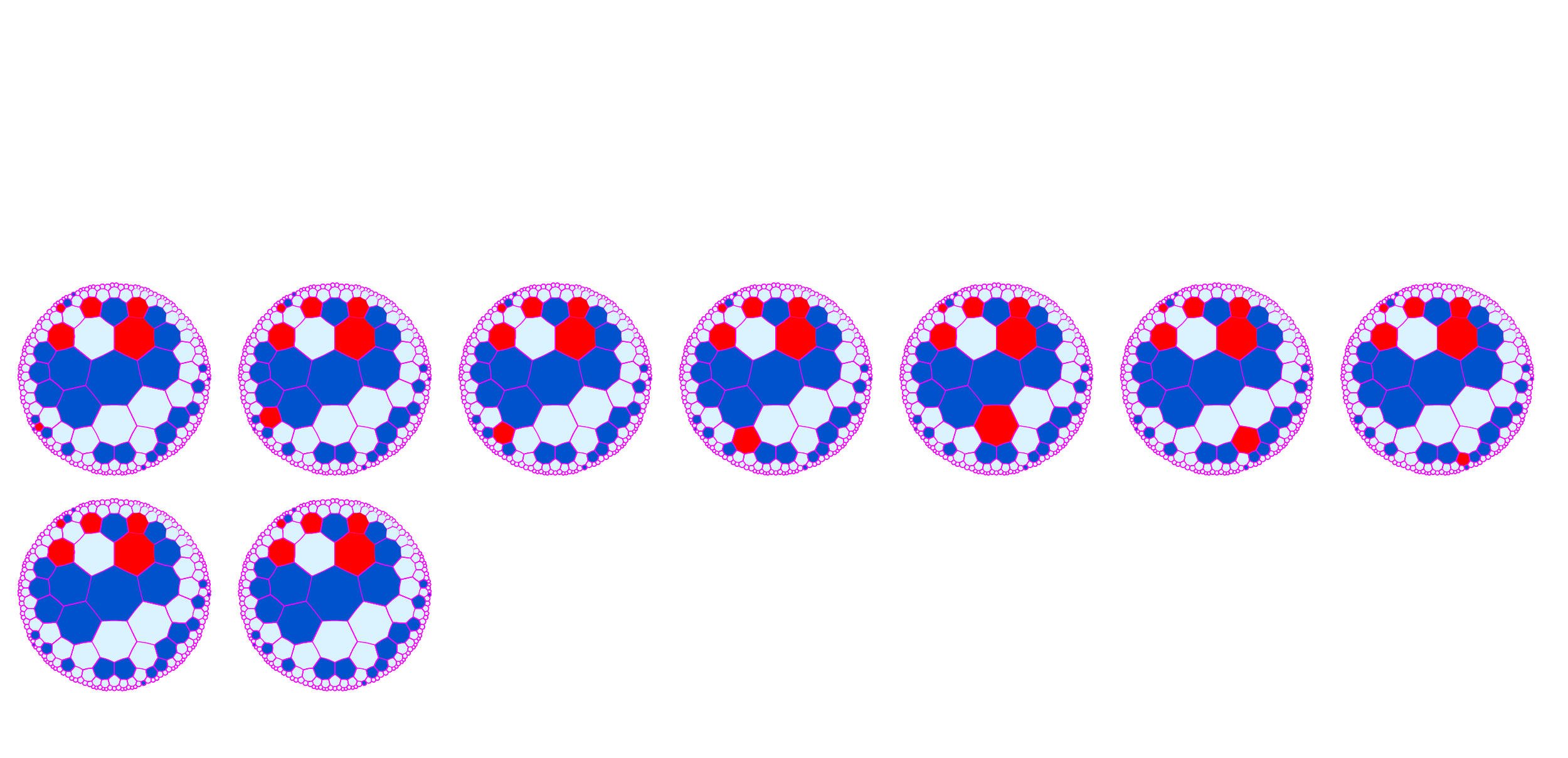}
\hfill}
\vspace{-70pt}
\ligne{\hfill
\includegraphics[scale=0.27]{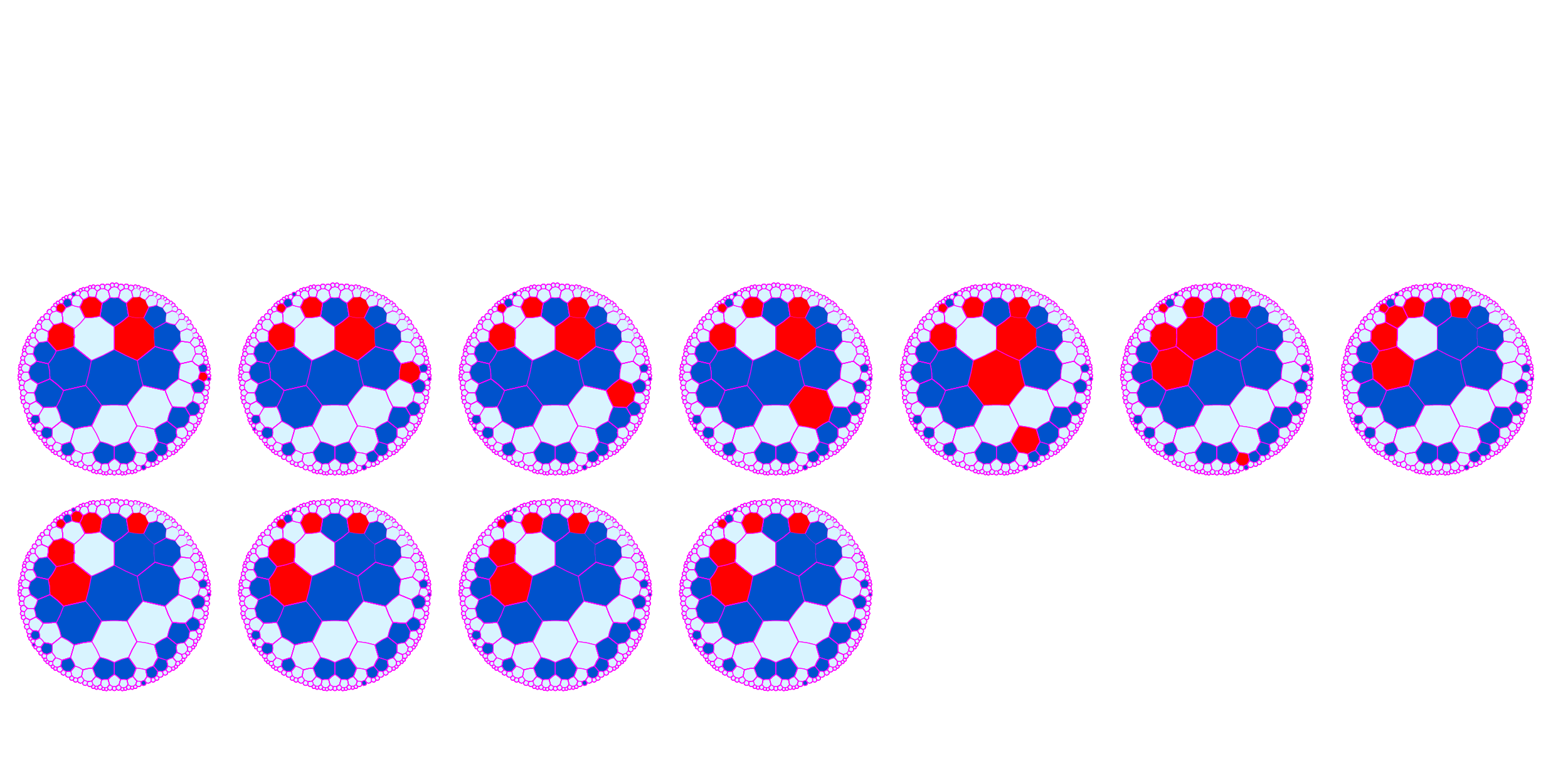}
\hfill}
\vspace{-70pt}
\ligne{\hfill
\includegraphics[scale=0.27]{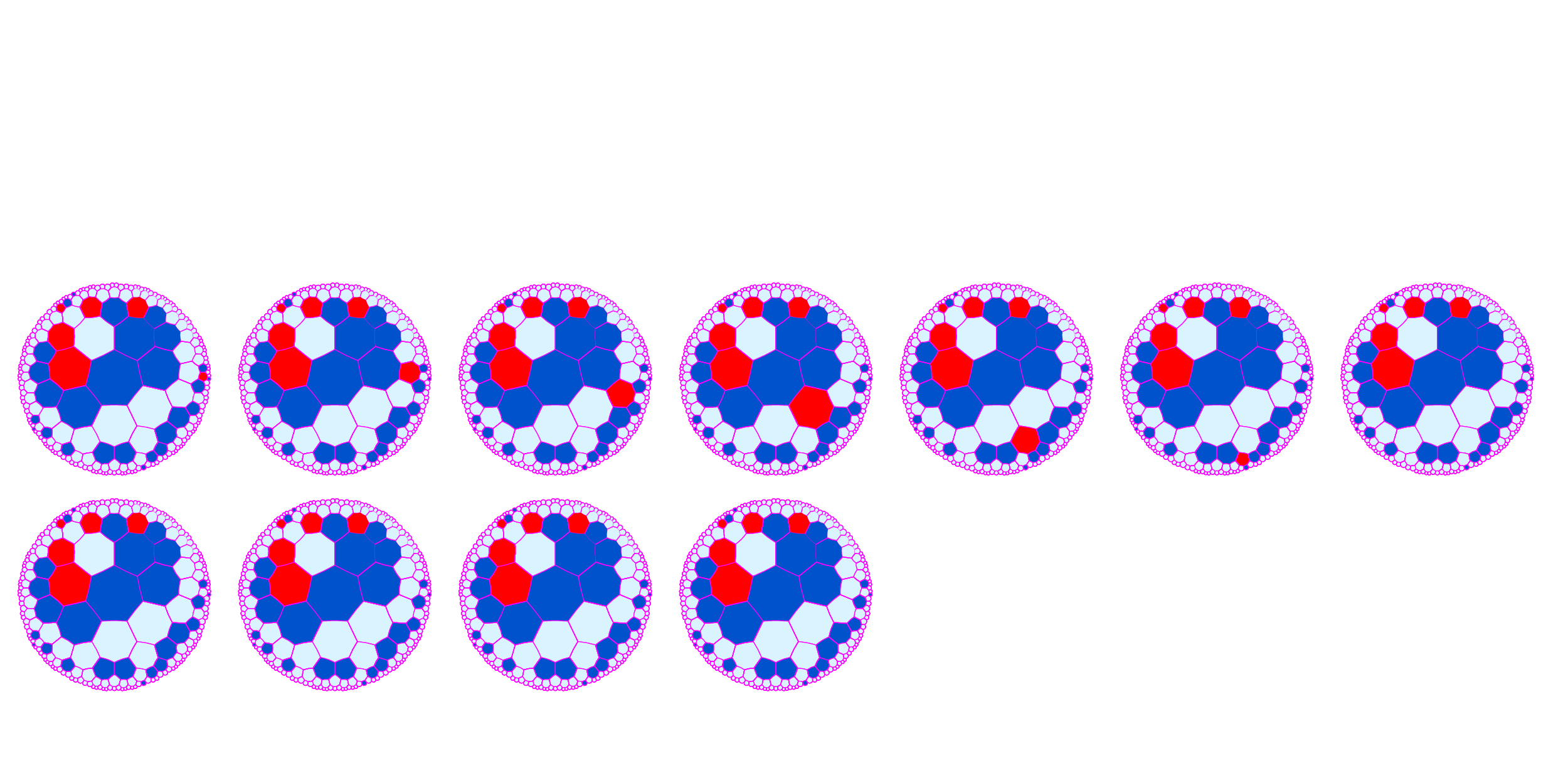}
\hfill}
\vspace{-70pt}
\ligne{\hfill
\includegraphics[scale=0.27]{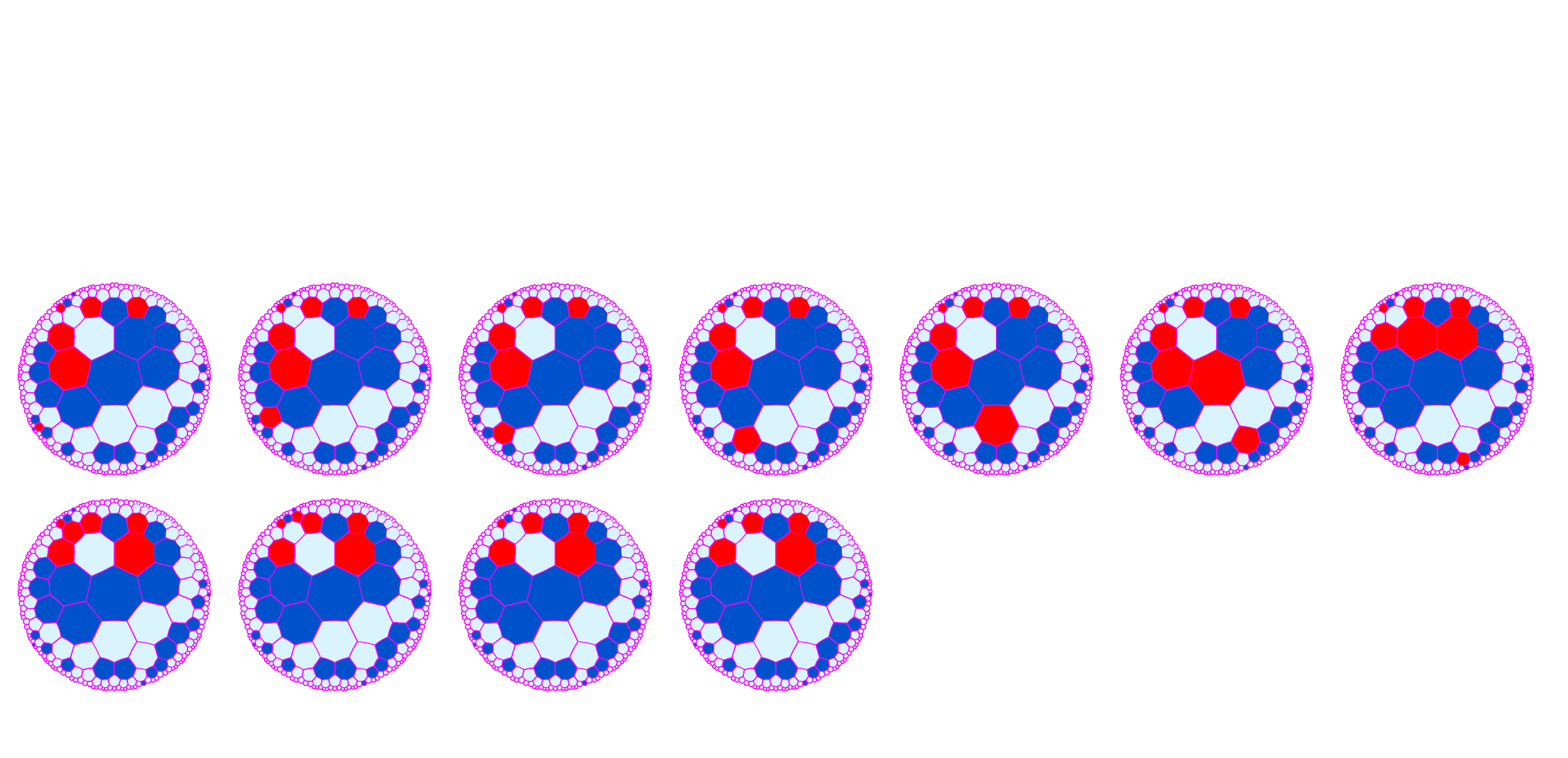}
\hfill}
\vspace{-15pt}
\begin{fig}\label{hca73memopassif}
\leurre
The passive memory switch. The two versions of the switch and the two possible
crossings: above, first two rows, left-hand side switch; below, last two rows, right-hand side one. 
\end{fig}
\vspace{10pt}
}

In the lower half, the figure illustrates the working of the right-hand side switch. In the first
two lines, the locomotive arrives from the right so that nothing is changed. In the last two lines
it arrives from the left: accordingly, the selection is changed to the left-hand side one.

\vskip 7pt
\noindent
\ligne{$\underline{\hbox{Implementation of the flip-flop and of the active memory switch}}$}
\vskip 5pt

   According to Figures~\ref{memoact_73} and~\ref{bascul_73} what we have to implement is 
the fork and the killer.

   First, let us have a look at Figure~\ref{hca73kill}. It illustrates the working of the killer.
In the upper half, we have the situtation of a locomotive arriving to the killer. We say
that the killer is blue, red depending on whether its central cell in Figure~\ref{hca73kill}
is blue or red.

\vtop{
\vspace{-45pt}
\ligne{\hfill
\includegraphics[scale=0.27]{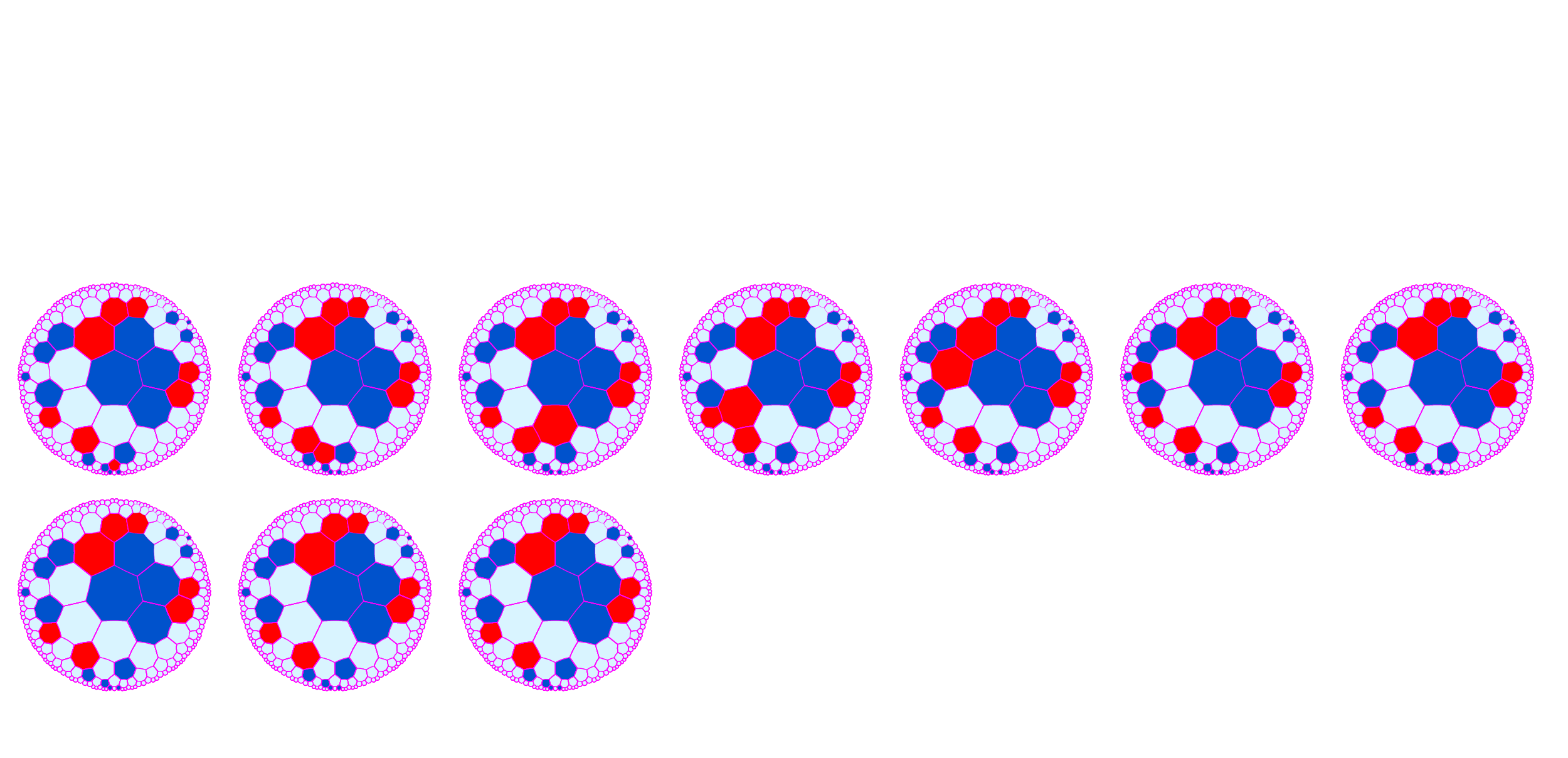}
\hfill}
\vspace{-70pt}
\ligne{\hfill
\includegraphics[scale=0.27]{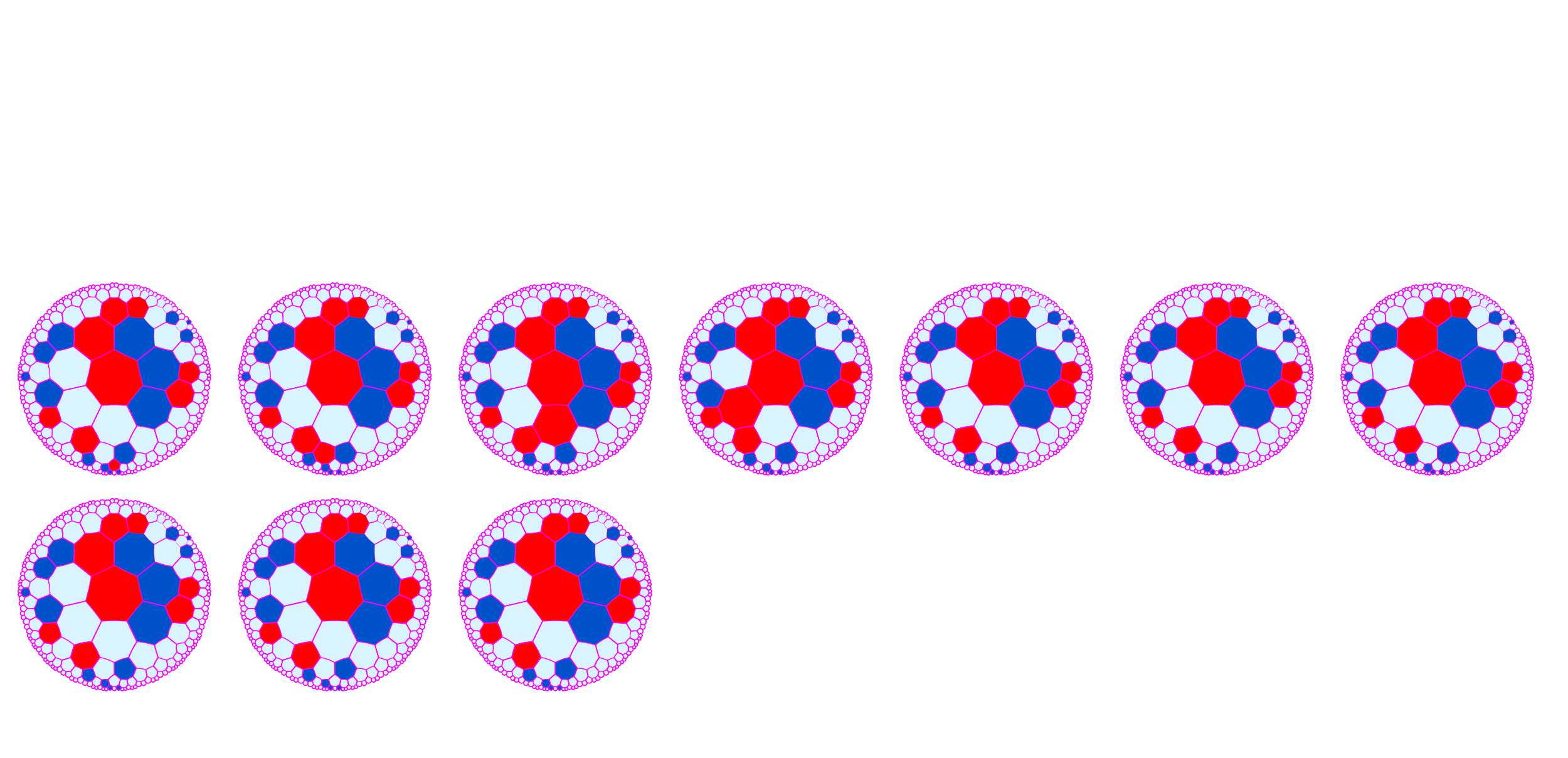}
\hfill}
\vspace{-70pt}
\ligne{\hfill
\includegraphics[scale=0.27]{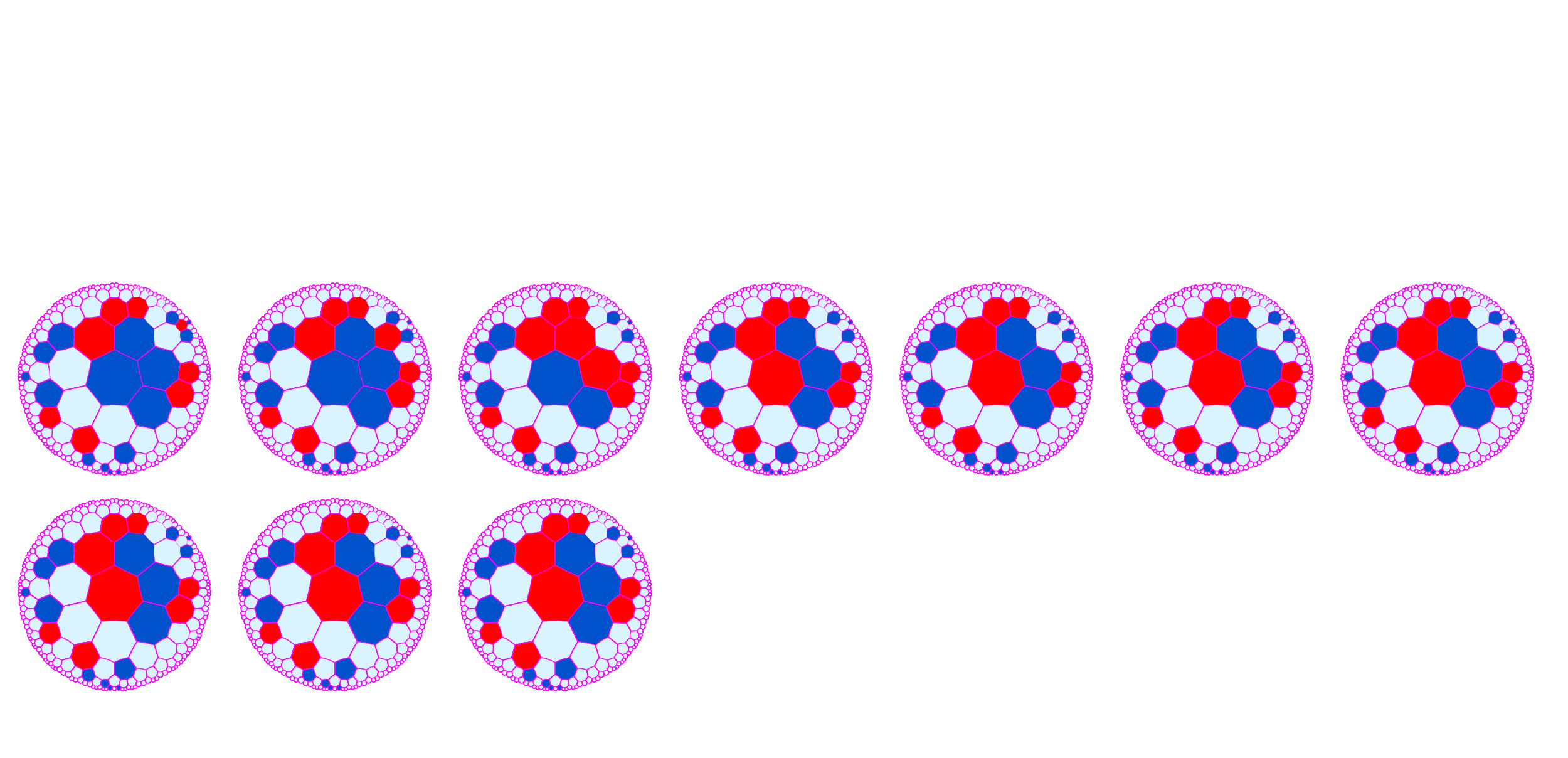}
\hfill}
\vspace{-70pt}
\ligne{\hfill
\includegraphics[scale=0.27]{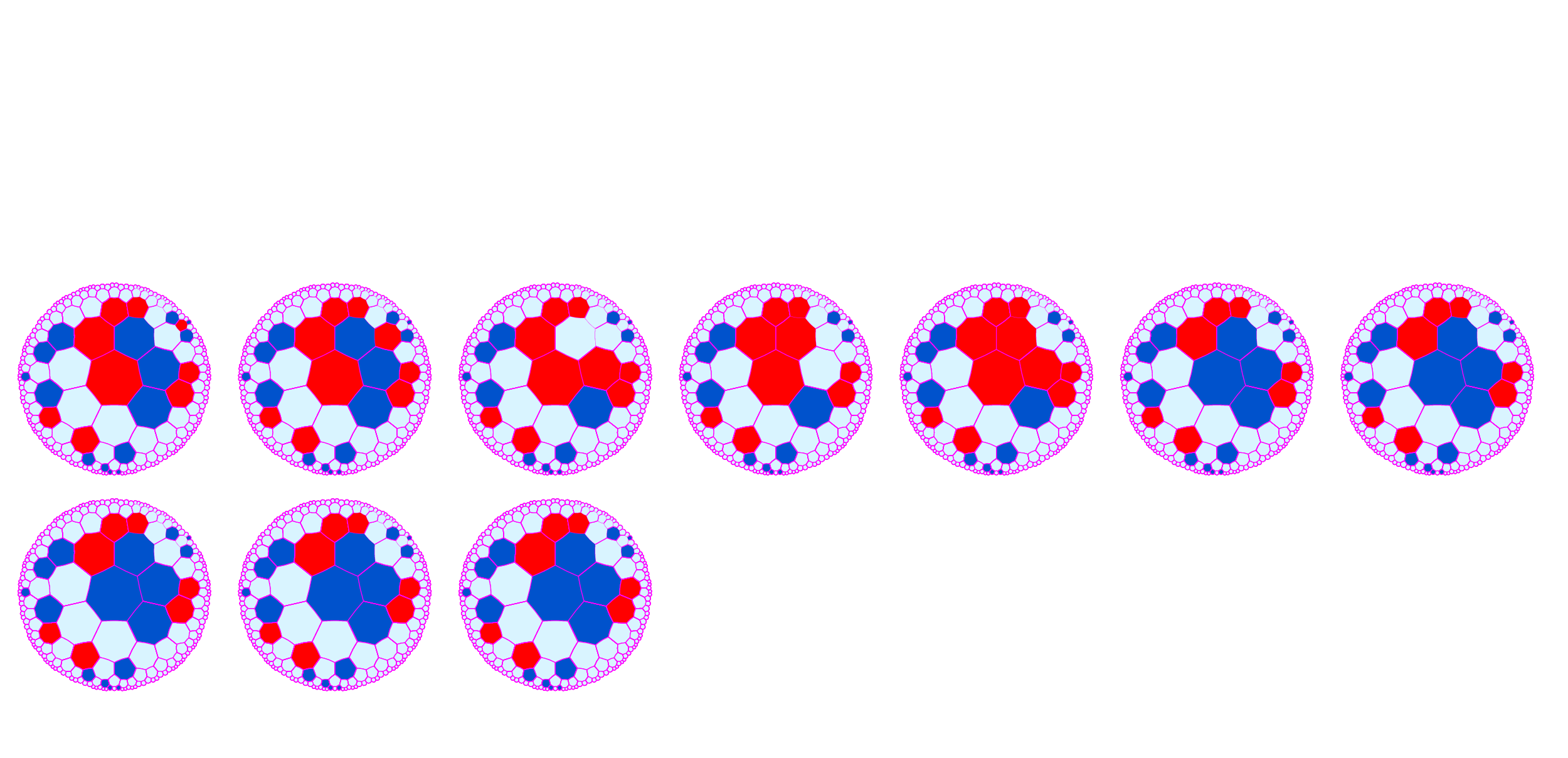}
\hfill}
\vspace{-15pt}
\begin{fig}\label{hca73kill}
\leurre
The control, structure used both by the flip-flop and by the active memory switch.
When killer is blue, the locomotive crosses the killer and goes on its way on the track.
When the killer is red, the locomotive is destroyed. The lower half of the figure: when
the signal arrives at the killer, it changes its colour.
\end{fig}
\vspace{10pt}
}

   On the upper half of the figure, we can see that if the killer is red, a locomotive arriving 
there cannot pass the killer. But when it is blue, the locomotive can pass the killer and so it
goes on its way on the track. The lower part of the figure shows us how a signal arriving to the
killer changes its colour. We can see that the change from red to blue takes 5~steps while
the change from blue to red takes 3~steps only. As will be seen in Section~\ref{rules},
in the change from red to blue, we have to take into accounts rules for other motions which
apply also in this case.
\vskip 5pt
   Figure~\ref{hca73fork} illustrates the fork. We can see that a locomotive arriving at the
fork is duplicated and each new locomotive goes on a different track.

\vtop{
\vspace{-45pt}
\ligne{\hfill
\includegraphics[scale=0.27]{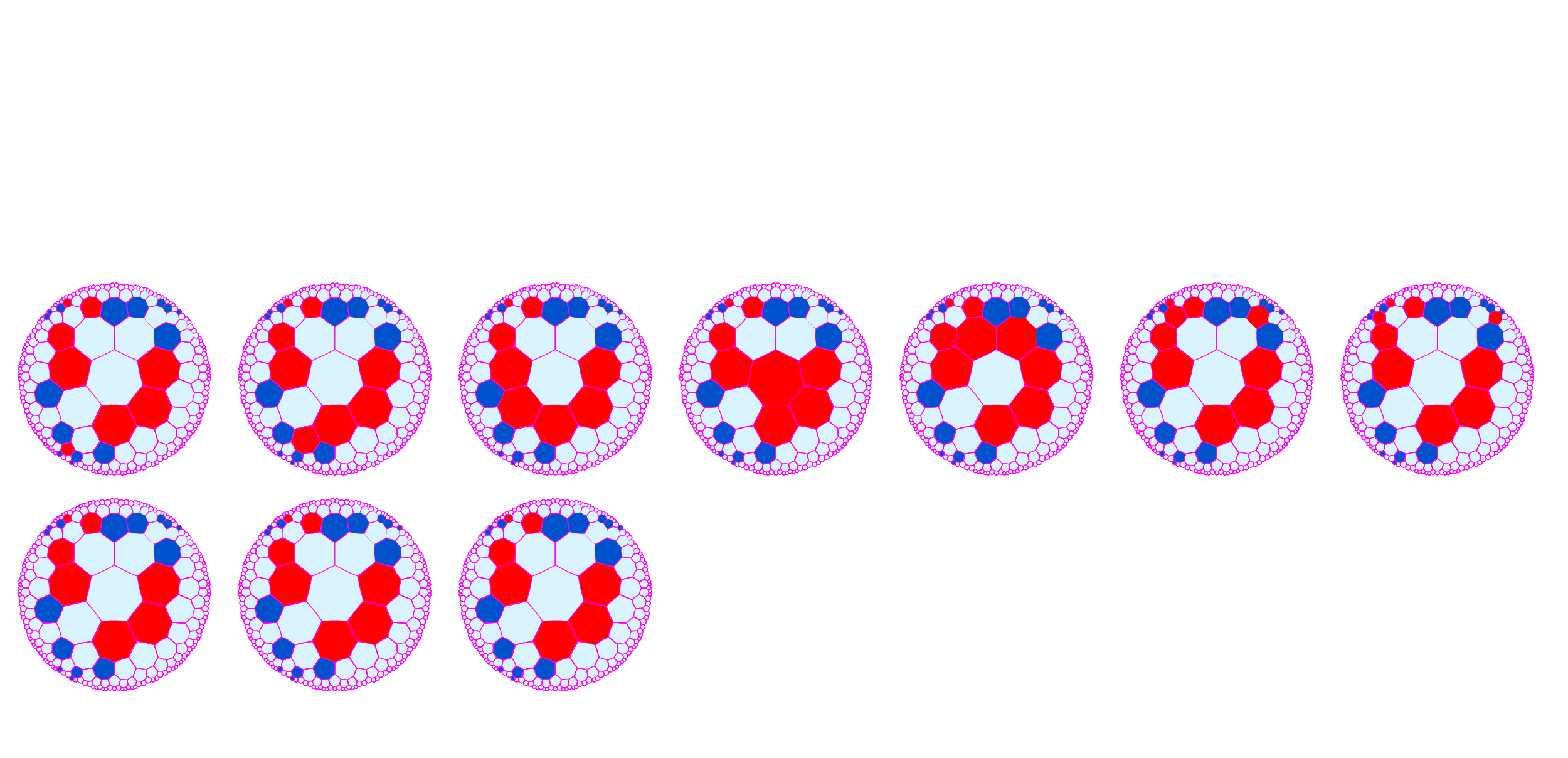}
\hfill}
\vspace{-15pt}
\begin{fig}\label{hca73fork}
\leurre
The fork, the other structure used both by the flip-flop and by the active memory switch.
\end{fig}
\vspace{10pt}
}

    In Sub-section~\ref{newscenar}, we have seen how the killer and the fork are combined in
order to obtain the working of a flip-flop or that of an active memory switch.

    It remains to remark that the signal wich arrives to both killers come from the switch itself
in the case of the flip-flop and that it comes from another switch in the case of the active memory
switch. In that latter case, the signal comes from the passive memory switch: when the locomotive
arrives from the non-selected track, Figure~\ref{hca73memopassif} shows that a signal is emitted
from the switch: see the lines~3 and~7 of the figure. The track which arrives to~$S$ in 
Figure~\ref{memoact_73} comes from a passive memory switch.

   Figure~\ref{newconnect} illustrates the connection between the passive memory switch, the blue
disk with a purple circle on the right-hand side of the figure, and the active memory switch,
the orange square, with a black border, on the left-hand side of the picture. We may imagine
that inside the orange square we have a copy of Figure~\ref{memoact_73}, considering that
the orange arrow which arrives to the square from above arrives to the fork~$S$ inside the
square. Also, the tracks which leave the killers in the copy of Figure~\ref{memoact_73} are the 
tracks which leave the upper corners of the orange square in Figure~\ref{newconnect}. With this 
figure, Figures~\ref{memoact_73} and~\ref{hca73kill}, we can check that the working of
the active memory switch is conformal to what we have described in Sub-section~\ref{newscenar}.

    Here, we may again note that up to three locomotives may occur at the same time on the 
circuit. Say that the time of execution of an elementary action in our circuit coincide with
the arrival of the locomotive at the central cell of either a rhombic pattern,
a passive memory switch, or a fork~$C$. Then, as already mentioned, we may construct the circuit in
such a way that when the locomotive starts an elementary action, it is the single locomotive of
the circuit. With this condition, it is clear that the just considered implementation
is correct with respect to the scenario described in Sub-section~\ref{newscenar}. 

\vtop{
\vspace{-5pt}
\ligne{\hfill
\includegraphics[scale=0.45]{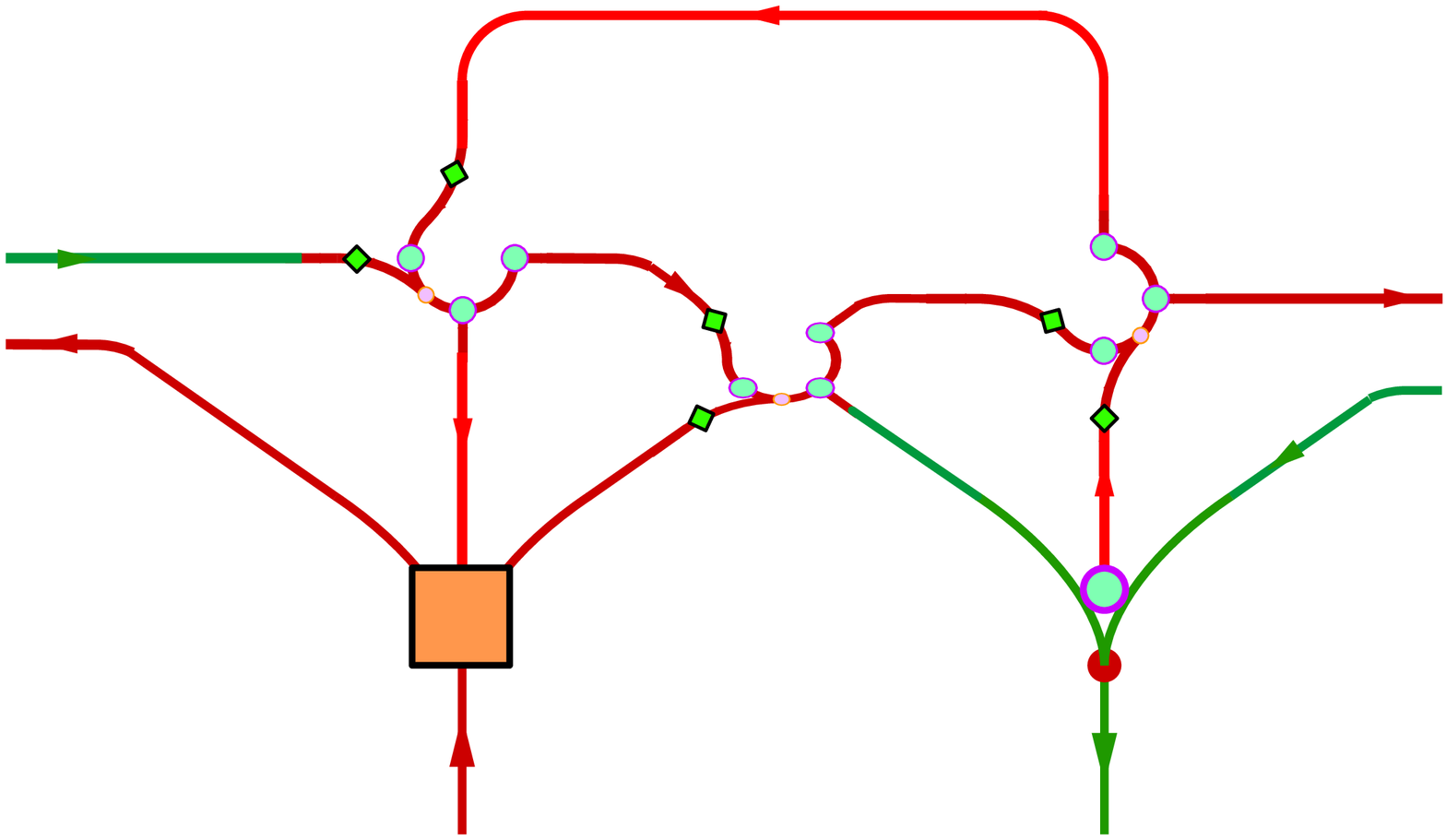}
\hfill}
\vspace{-5pt}
\begin{fig}\label{newconnect}
\leurre
The connection from the passive memory switch to the active one. The active switch is
represented by the big orange square.
\end{fig}
\vspace{10pt}
}

\section{The rules}
\label{rules}

   In order to prove the existence of the cellular automaton whose working was
descrived in the previous sections, we have to implement its rules. Here, we display 
all of them, but we list them in several groups according to the presentation of the previous
implementation. 
We present the rules according to the same order as we presented the implementation in
Subsection~\ref{scenar}

   The rules are {\bf rotation invariant}.
   This means that if we perform a circular permutation on the neighbours of a cell, this
does not change the new state. Accordingly, in the tables we give, the rules are rotationally
invariant: a given rule~$\rho$ has four over rules with the same current state and the same 
new state, the states of the neighbours being a circular permutation of those indicated by~$\rho$.
 The rules have been checked by a computer program
written by the author. The program insures that the rules are rotation invariant and that
there is not conflict between them. Also, the program construced the figures used in
Section~\ref{implement}, so that it allowed me to check that the rules make the cellular automaton
to perform what it is expected to do.

\subsubsection{Rules for the tracks}

   First, we have the rules for the motion of the locomotive on the tracks which are displayed
in Table~\ref{rultrack}. 
   The rules are presented in the same order as established while checking them during the 
construction by a computer program.

   As indicated in the caption of the table, the first two columns of rules are used for the
vertical tracks and the last two columns contain additional rules needed by the
horizontal tracks. Of course, as the standard element is present in both kinds of tracks,
several rules used for the vertical tracks are also used for the horizontal ones.
Some rules are {\bf conservative}: they do not change the state of the cell. In particular,
these rules apply to those of the milestones which remain unchanged while the locomotive
passes close to them. However, some of them may change their state for one or two steps of the
computation. Then we apply the other rules which are not conservative. Many rules are {\bf motion}
rules: either they apply to the locomotive itself or two both of them when two of them are 
contiguous. We also consider as motion rules, rules which change the state of a milestone
in order to contribute to the motion. This is the case, in particular, for the rules which
apply to those {\bf signals} which may change, according to the kind of switch to which they are
associated. Note that in many cases a motion rule is not conservative, but in a few cases, a
motion rule is also conservative: in particular, when the locmotive leaves a cell~$c$, the
rules which apply to~$c$ just after the locomotive left it is conservative.

As an example of illustration of motion rules, consider rules~7 up to~11. Rules~7, 8, 9 and~11 
givern the motion of a single locomotive as it can easily be checked on figure~\ref{elemvoie}.
Rule~10 is an additional rule needed by the motion of two contiguous locomotives moving together.

\newcount\numero\numero=0
\def\arule #1 #2 #3 #4 #5 #6 #7 #8 #9 {%
\footnotesize\tt
\ligne{\global\advance \numero by 1
   \hbox to 20pt {\hfill\the\numero\hskip 5pt}
\hbox to 57pt {%
$\underline{\hbox{\tt #1}}$#2#3#4#5#6#7#8$\underline{\hbox{\tt #9}}$\hfill}
\hfill}
\vskip 0pt
}

\newdimen\largoo\largoo=80pt

\setbox110=\vtop{\leftskip 0pt\parindent 0pt\hsize=\largoo
\baselineskip 10pt
\arule  W   W   W   W   W   W   W   W   W   
\arule  B   W   W   W   W   W   W   W   B   
\arule  B   B   W   W   W   W   W   W   B   
\arule  W   B   W   W   W   W   W   W   W   
\arule  W   B   B   W   W   W   W   W   W   
\arule  W   B   B   W   B   W   W   W   W   
\arule  W   W   B   W   B   W   B   W   W   
\arule  W   W   B   W   B   R   B   W   R   
\arule  R   W   B   W   B   W   B   W   W   
\arule  R   W   B   R   B   W   B   W   W   
\arule  W   W   B   R   B   W   B   W   W   
\arule  R   W   B   W   B   R   B   W   R   
\arule  B   R   W   W   W   W   W   W   B   
\arule  W   B   R   B   W   W   W   W   W   
\arule  W   B   R   W   W   W   W   W   W   
\arule  W   R   B   W   W   W   W   W   W   
\arule  W   B   W   B   W   W   W   W   W   
\arule  B   B   W   R   W   W   W   W   B   
}

\setbox112=\vtop{\leftskip 0pt\parindent 0pt\hsize=\largoo
\baselineskip 10pt
\arule  B   B   W   W   R   W   W   W   B   
\arule  W   B   B   W   B   W   B   W   W   
\arule  W   B   B   W   B   R   B   W   R   
\arule  R   B   B   W   B   W   B   W   W   
\arule  W   B   B   W   B   W   B   R   W   
\arule  R   B   B   W   B   R   B   W   R   
\arule  R   B   B   W   B   W   B   R   W   
\arule  B   B   R   W   W   W   W   W   B   
\arule  B   R   B   W   W   W   W   W   B   
\arule  W   R   B   W   W   W   B   B   W   
\arule  B   R   W   B   W   W   W   W   B   
\arule  B   W   W   B   W   W   W   R   B   
\ligne{\hfill}
}

\setbox114=\vtop{\leftskip 0pt\parindent 0pt\hsize=\largoo
\baselineskip 10pt
\arule  W   B   B   W   B   B   B   W   W   
\arule  W   B   B   R   B   B   B   W   R   
\arule  R   B   B   W   B   B   B   W   W   
\arule  W   B   B   W   B   B   B   R   W   
\arule  R   B   B   R   B   B   B   W   R   
\arule  R   B   B   W   B   B   B   R   W   
\arule  B   B   B   W   W   W   W   W   B   
\arule  B   B   B   B   W   W   W   W   B   
\arule  B   B   W   B   W   W   W   W   B   
\arule  B   B   B   R   W   W   W   W   B   
\arule  B   B   B   W   R   W   W   W   B   
\arule  B   B   B   W   W   R   W   W   B   
\arule  B   B   B   W   W   W   R   W   B   
\arule  B   B   B   W   W   W   W   R   B   
\arule  B   R   B   B   B   W   W   W   B   
\arule  B   B   R   B   W   W   W   W   B   
\arule  B   B   B   R   W   W   W   B   B   
\arule  B   B   B   W   R   W   W   B   B   
}

\setbox116=\vtop{\leftskip 0pt\parindent 0pt\hsize=\largoo
\baselineskip 10pt
\arule  B   B   B   W   W   R   W   B   B   
\arule  B   R   R   W   B   W   W   W   B   
\arule  B   R   R   W   W   W   W   W   B   
\arule  B   B   R   R   W   W   W   W   B   
\arule  B   B   W   R   R   W   W   W   B   
\arule  B   B   W   W   R   R   W   W   B   
\arule  B   B   B   R   R   W   W   W   B   
\arule  B   B   B   W   R   R   W   W   B   
\arule  B   B   B   W   W   R   R   W   B   
\arule  B   B   B   W   W   W   R   R   B   
\arule  B   R   R   B   W   W   W   W   B   
\arule  B   B   B   B   R   R   W   W   B   
\arule  B   B   B   B   W   R   R   W   B   
\arule  B   B   B   B   W   W   R   R   B   
}

\vtop{
\vspace{-10pt}
\begin{tab}\label{rultrack}
\leurre
The rules for the tracks. The first two columns are rules for the vertical tracks.
The last twocolumns are additonal rules required by the horizontal ones.
\end{tab}
\vspace{-10pt}
\ligne{\hfill\box110\hfill\box112\hfill\box114\hfill\box116\hfill}
}
\vskip 7pt
  And so, we can see that the motion on the tracks requires 62~rules.

\subsubsection{Rules for the crossing}

   Table~\ref{rulcross} provides additional rules for the crossing. It can be noticed
in Figures~\ref{hca73croisdb} and~\ref{hca73croisav} that the central cell of the rhombic
pattern and pattern~{\bf 1} have different neighbourhoods. For the rhombic pattern
it is {\tt\footnotesize BBWRRRW} and for pattern~{\bf 1} it is 
{\tt\footnotesize BBBWRRW}. The \WW-neighbours of the \BB-part of the ring around the central
cell is the same in both patterns, but they are placed in opposite places which explains
the difference of working of the patterns and also the fact that they share many rules
so that it is not easy to clearly identify which rules are used in one of these patterns and not
in the other.

\setbox110=\vtop{\leftskip 0pt\parindent 0pt\hsize=\largoo
\baselineskip 10pt
\arule  W   B   B   B   W   R   R   W   W   
\arule  W   B   B   B   W   R   R   R   W   
\arule  W   B   B   B   W   R   W   W   R   
\arule  R   B   B   B   W   R   R   W   W   
\arule  W   B   B   B   R   R   R   W   W   
\arule  W   W   R   W   B   W   B   W   W   
\arule  W   W   R   W   B   R   B   W   R   
\arule  R   W   R   W   B   W   B   W   W   
\arule  W   W   R   R   B   W   B   W   W   
\arule  W   W   B   W   R   W   B   W   W   
\arule  W   W   B   W   R   R   B   W   R   
\arule  R   W   B   W   R   W   B   W   W   
\arule  R   W   B   W   R   R   B   W   R   
\arule  R   W   B   R   R   W   B   W   R   
\arule  W   W   B   R   R   W   B   W   W   
\arule  R   B   R   W   W   W   W   W   R   
\arule  R   B   W   W   W   W   W   W   R   
\arule  R   B   R   R   W   W   W   W   R   
}

\setbox112=\vtop{\leftskip 0pt\parindent 0pt\hsize=\largoo
\baselineskip 10pt
\arule  R   B   R   W   R   W   W   W   R   
\arule  R   B   R   W   W   R   W   W   R   
\arule  R   R   B   W   W   W   W   W   R   
\arule  W   B   B   W   R   W   B   W   W   
\arule  W   B   B   W   R   R   B   W   R   
\arule  R   B   B   W   R   W   B   W   W   
\arule  W   B   B   W   R   W   B   R   W   
\arule  R   B   B   W   R   R   B   W   R   
\arule  R   B   B   W   R   W   B   R   W   
\arule  W   R   W   B   W   W   W   W   W   
\arule  W   R   R   B   W   W   W   W   W   
\arule  W   R   B   W   B   W   B   W   W   
\arule  W   R   B   W   B   R   B   W   R   
\arule  R   R   B   W   B   W   B   W   W   
\arule  W   R   B   W   B   W   B   R   W   
\arule  R   R   B   W   B   R   B   W   R   
\arule  R   R   B   W   B   W   B   R   W   
\arule  W   R   W   W   W   W   W   W   W   
}

\setbox114=\vtop{\leftskip 0pt\parindent 0pt\hsize=\largoo
\baselineskip 10pt
\arule  R   R   B   R   W   W   W   W   R   
\arule  R   R   B   W   R   W   W   W   R   
\arule  R   R   B   W   W   R   W   W   R   
\arule  R   R   B   W   W   W   R   W   R   
\arule  R   R   B   W   W   W   W   R   R   
\arule  R   R   B   R   R   W   W   W   W   
\arule  W   R   B   W   R   R   W   W   R   
\arule  B   R   R   R   W   W   W   W   B   
\arule  W   W   B   W   B   W   B   R   W   
\arule  R   W   W   B   W   W   B   R   W   
\arule  W   W   W   R   B   W   B   B   W   
\arule  W   R   R   W   B   W   B   B   W   
\arule  R   W   W   R   B   W   W   B   W   
\arule  B   B   W   W   B   W   W   W   B   
\arule  B   B   W   R   B   W   W   W   B   
\arule  B   B   R   W   B   W   W   W   B   
\arule  W   R   R   R   W   B   B   W   W    
\arule  W   R   R   R   R   B   B   W   R    
}

\setbox116=\vtop{\leftskip 0pt\parindent 0pt\hsize=\largoo
\baselineskip 10pt
\arule  R   R   R   R   W   B   B   W   R    
\arule  R   R   B   R   W   B   B   R   W    
\arule  W   R   R   R   W   B   B   R   W    
\arule  R   R   B   B   W   W   W   W   R    
\arule  W   B   W   B   B   W   W   W   W    
\arule  W   B   R   B   B   W   W   W   W    
\arule  R   R   R   W   W   W   B   B   R    
\arule  R   W   B   R   W   R   B   W   R    
\arule  R   W   B   R   R   R   B   W   B    
\arule  B   W   B   R   R   R   B   W   R    
}

\vtop{
\vspace{-5pt}
\begin{tab}\label{rulcross}
\leurre
The rules for the crossings.
\end{tab}
\vspace{-10pt}
\ligne{\hfill\box110\hfill\box112\hfill\box114\hfill\box116\hfill}
}

\vskip 7pt
We can see from Tables~\ref{rultrack} and~\ref{rulcross} that the crossings require
64~additional rules, not taking into account the many rules from Table~\ref{rultrack} it uses too.

\subsubsection{Rules for the fixed switch}

   These rules are displayed by Table~\ref{rulfix}.
 
\setbox110=\vtop{\leftskip 0pt\parindent 0pt\hsize=\largoo
\baselineskip 10pt
\arule  W   W   B   W   B   B   B   W   W    
\arule  W   R   B   W   B   B   B   W   R    
\arule  R   W   B   W   B   B   B   W   W    
\arule  W   W   B   R   B   B   B   W   W    
\arule  W   W   W   B   B   W   B   B   W    
\arule  W   W   W   B   B   R   B   B   R    
\arule  R   W   W   B   B   W   B   B   W    
\arule  B   W   B   W   W   B   R   W   B    
\arule  B   W   B   W   W   B   W   R   B    
\arule  B   W   B   W   W   B   R   R   B    
\arule  B   R   B   W   W   B   W   R   B    
\arule  B   R   B   W   W   B   W   W   B    
\arule  W   W   B   B   W   B   B   R   W    
\arule  W   R   B   B   W   B   B   R   W    
\arule  W   R   B   B   W   B   B   W   W    
}

\setbox112=\vtop{\leftskip 0pt\parindent 0pt\hsize=\largoo
\baselineskip 10pt
\ligne{--\hfill} 
\arule  W   W   B   W   B   B   B   R   R    
\arule  R   R   B   B   W   B   B   W   W    
\arule  B   R   B   W   W   W   B   R   B    
\ligne{\ \ \tt\footnotesize deux \hfill} 
\ligne{\ \ \tt\footnotesize locomotives \hfill} 
\arule  R   R   B   W   B   B   B   W   R    
\arule  R   W   B   R   B   B   B   W   W    
\arule  R   W   W   B   B   R   B   B   R    
\arule  R   R   W   B   B   W   B   B   W    
\ligne{--\hfill} 
\arule  R   W   B   W   B   B   B   R   R 
}

\setbox114=\vtop{\leftskip 0pt\parindent 0pt\hsize=\largoo
\baselineskip 10pt
}
\vtop{
\vspace{-5pt}
\begin{tab}\label{rulfix}
\leurre
The rules for the fixed switch. 
\end{tab}
\vspace{-10pt}
\ligne{\hfill\box110\hfill\box112\hfill\box114\hfill\box116\hfill}
}

   As mentioned in the table, the switch may be crossed either by a single locomotive
or by two locmotives always running contiguously.

\subsubsection{Rules for the passive memory switch}

   Table~\ref{rulmemopass} gives the rules for the passive memory switch. It concerns both
versions of the switch: one when the left-hand side track is selected, the other when the right-hand
side track is selected. The cells which signalize the selected and the non-selected tracks 
are the cells sharing side~2 and side~7 of the central cell which is a blue milestone,
see Figure~\ref{hca73memopassif}. In the first two rows of the figure, the red signal is in
the cell sharing side~7.

\setbox110=\vtop{\leftskip 0pt\parindent 0pt\hsize=\largoo
\baselineskip 10pt
\ligne{\ \ \tt\footnotesize gauche, venue\hfill}
\ligne{\ \ \tt\footnotesize de droite\hfill}
\arule  B   W   B   B   W   W   B   R   B    
\arule  B   W   B   B   W   R   B   R   R    
\arule  R   W   B   B   W   W   B   R   B    
\arule  B   R   R   B   W   W   B   B   B    
\arule  B   W   R   B   W   W   B   B   B    
\arule  W   B   R   B   R   W   R   B   W    
\arule  W   R   R   B   R   W   R   B   R    
\arule  R   B   B   B   R   W   R   R   W    
\arule  W   B   B   B   R   W   R   R   W    
\arule  W   B   B   B   R   R   R   R   W    
\arule  B   W   R   R   W   W   W   R   B    
\arule  B   R   B   R   W   W   W   R   B    
\arule  B   W   B   R   W   W   W   R   B    
\arule  B   W   R   B   W   W   W   R   B    
\arule  R   W   B   W   W   W   W   R   R    
\arule  R   W   B   W   W   W   R   W   R    
\arule  W   W   R   W   B   R   W   R   W    
\arule  W   R   R   W   B   R   W   R   R    
}

\setbox112=\vtop{\leftskip 0pt\parindent 0pt\hsize=\largoo
\baselineskip 10pt
\arule  R   W   R   W   B   R   W   R   W    
\arule  W   W   R   R   B   R   W   R   W    
\arule  W   R   W   W   W   W   W   R   W    
\arule  W   W   R   W   W   B   W   B   W    
\arule  W   R   R   W   W   B   W   B   R    
\arule  R   W   R   W   W   B   W   B   W    
\arule  B   W   R   W   W   W   W   R   B    
\arule  B   B   W   R   B   B   B   B   B    
\arule  B   R   W   R   B   B   B   B   R    
\arule  R   B   R   R   B   B   B   B   R    
\arule  R   B   W   R   B   B   B   B   R    
\arule  R   B   W   W   W   W   W   B   R    
\arule  W   R   W   R   W   W   W   W   W    
\arule  W   R   R   R   W   W   W   W   W    
\arule  W   R   B   W   B   B   R   W   W    
\arule  W   R   W   R   B   B   W   B   W    
\arule  B   B   W   W   R   W   B   R   B    
\arule  B   B   W   R   W   W   B   R   B    
}

\setbox114=\vtop{\leftskip 0pt\parindent 0pt\hsize=\largoo
\baselineskip 10pt
\arule  B   B   R   W   W   W   B   R   B    
\arule  W   B   R   W   W   W   W   B   W    
\arule  R   B   B   B   B   R   B   W   R    
\arule  R   R   B   B   B   R   B   W   B    
\arule  B   B   B   B   B   R   B   R   B    
\arule  B   B   B   B   B   R   B   W   B    
\arule  R   R   B   W   W   W   W   B   R    
\arule  R   B   B   W   W   W   W   B   R    
\ligne{\ \ \tt\footnotesize droite, venue\hfill}
\ligne{\ \ \tt\footnotesize de gauche\hfill}
\arule  R   W   R   B   W   W   B   B   B    
\arule  B   R   B   B   W   W   B   R   B    
\arule  W   R   B   B   R   W   R   R   R    
\arule  R   B   R   B   R   W   R   B   W    
\arule  W   B   R   B   R   R   R   B   W    
\arule  R   W   R   W   B   R   W   B   W    
\arule  B   R   R   R   W   W   W   R   B    
\arule  B   B   R   R   B   B   B   B   B    
\arule  R   B   R   W   W   W   W   B   R    
}

\setbox116=\vtop{\leftskip 0pt\parindent 0pt\hsize=\largoo
\baselineskip 10pt
\arule  R   B   W   R   W   W   W   B   R    
\arule  R   R   W   R   B   B   B   B   B    
\arule  B   R   B   B   B   R   B   W   R    
\arule  R   B   B   B   B   R   B   R   R    
\arule  B   B   R   B   W   W   W   R   B    
\ligne{\ \ \tt\footnotesize gauche, venue\hfill}
\ligne{\ \ \tt\footnotesize de gauche\hfill}
\arule  B   W   B   B   R   W   B   R   B
\ligne{\ \ \tt\footnotesize droite, venue\hfill}
\ligne{\ \ \tt\footnotesize de droite\hfill}
\arule  B   W   R   B   W   R   B   B   B
}

\vtop{
\vspace{-5pt}
\begin{tab}\label{rulmemopass}
\leurre
The rules for the passive memory switch.
\end{tab}
\vspace{-10pt}
\ligne{\hfill\box110\hfill\box112\hfill\box114\hfill\box116\hfill}
\vskip 10pt
}

   We note that when the locomotive, always a single one, crosses the switch through the non-selected
track, the central cell turns from blue to red for just one time. This is performed by rule~150
when the locomotive comes from the right-hand side and rule~195 when it comes from the left-hand 
side. The shutdown of this signal is performed by rules~152 and~195 repsectively. We have then 
rules which perform the exchange between red and blue in both neighbours
of the central cell sharing side~2 and side~7: rules~189 and~205 for the cell sharing side~7,
rules~176 and~204 for the cell sharing side~1. At the same time, the cell sharing side~1 turns from
white to red for just this time: this triggers a new locomotive which is sent to the 
active switch. The rule initiating the locomotive is rule~156, and rule~157 turns back the
cell to white, contributing to the motion of the new locomotive.

\setbox110=\vtop{\leftskip 0pt\parindent 0pt\hsize=100pt
\obeylines
\obeyspaces\global\let =\ \footnotesize\tt
  2    1    7    0

  B    W    R    B 
  B    W    R    R 
  R    R    B    B 
  R    W    B    B 
}

\setbox112=\vtop{\leftskip 0pt\parindent 0pt\hsize=100pt
\obeylines
\obeyspaces\global\let =\ \footnotesize\tt
  2    1    7    0

  R    W    B    B 
  R    W    B    R 
  B    R    R    B 
  B    W    R    B 
}

\vtop{
\begin{tab}\label{basc_exec}
\leurre
Execution trace of the crossing of the passive memory switch by the locomotive through the 
non-selected track. The neighbours are identified by the side shared with the central cell.
\end{tab}
\vspace{-5pt}
\ligne{\hfill\box110\hfill\box112\hfill}
}

\subsubsection{Rules for the fork and for the killer}

   Here, the rules will be divided into two tables: Table~\ref{rulkill} and 
Table~\ref{rulfork} for the killer and for the fork respectively.

   As explained in Section~\ref{implement}, the killer is a controlling structure. In 
Figure~\ref{hca73kill}, the central cell is a milestone which is a signal for the locomotive.
Note that a single locomotive arrives to the killer. If the signal is blue, the locomotive
goes on its way and leaves the structure. If the signal is red, the locomotive is destroyed
by the structure. For instance, rules~217 and~218 which apply to the central cell
witness that the locomotive passes its way when the signal is blue. Rules~219, 220 and~221
apply when the signal is red and we can see on rules~219 and~221 that the locomotive disappears:
rule~219 is applied just after rule~221 was applied. The 'killing' rule is rule~168, a rule
already used by the passive memory switch.

\vskip 5pt
\setbox110=\vtop{\leftskip 0pt\parindent 0pt\hsize=\largoo
\baselineskip 10pt
\arule  W   B   R   W   R   W   B   W   W    
\arule  W   B   W   B   W   W   W   R   W    
\arule  W   B   R   W   R   R   B   W   R    
\arule  R   B   R   W   R   W   B   W   W    
\arule  W   B   R   W   R   W   B   R   W    
\arule  R   W   W   W   W   B   W   R   R    
\arule  W   R   R   W   B   W   B   R   W    
\ligne{--\hfill}
\arule  B   R   W   W   R   B   B   B   B    
\arule  B   R   W   R   W   B   B   B   B    
\arule  R   R   W   W   W   B   B   B   R    
\arule  R   R   W   W   R   B   B   B   R    
\arule  R   R   W   R   W   B   B   B   R    
}

\setbox112=\vtop{\leftskip 0pt\parindent 0pt\hsize=\largoo
\baselineskip 10pt
\arule  B   R   W   W   W   B   R   R   R    
\arule  R   R   W   W   W   B   R   R   B    
\arule  R   B   B   R   W   W   B   W   R    
\arule  R   B   B   R   W   W   B   R   R    
\arule  R   R   B   R   W   W   B   W   R    
\arule  R   B   R   R   W   W   B   W   R    
\arule  W   B   R   B   B   W   B   R   R    
\arule  W   B   R   B   B   R   B   W   W    
\arule  W   R   R   B   B   W   B   W   W    
\arule  W   R   R   B   B   W   B   R   W    
\arule  B   B   R   W   W   W   R   B   B    
\arule  B   R   W   W   W   W   R   B   B    
\arule  B   R   R   W   W   W   R   B   B    
\arule  B   B   B   R   R   W   R   B   R    
\arule  B   R   B   R   R   W   W   B   B    
\arule  R   B   B   R   R   W   W   R   B    
\arule  W   B   R   W   W   W   B   W   W    
\arule  W   B   R   W   W   W   B   R   W    
}

\setbox114=\vtop{\leftskip 0pt\parindent 0pt\hsize=\largoo
\baselineskip 10pt
\arule  W   R   R   W   W   W   B   W   W    
\arule  B   B   B   W   W   R   R   R   B    
\arule  B   R   B   W   W   R   R   R   B    
\arule  B   B   B   R   W   R   R   R   R    
\arule  B   R   B   R   W   R   R   R   W    
\arule  R   B   R   W   W   R   R   R   B    
\arule  W   R   R   W   B   W   B   W   W    
\arule  W   B   R   B   W   W   W   R   W    
\arule  W   R   W   B   W   W   W   R   W    
\ligne{--\hfill}
\arule  R   R   R   R   W   W   B   W   R    
\arule  B   R   B   R   R   W   R   B   R    
\arule  W   R   B   R   R   W   W   R   R    
\arule  R   R   B   R   R   W   W   R   B    
\arule  R   R   W   W   W   R   R   R   R    
\arule  R   R   R   W   W   R   R   R   B    
\arule  W   W   R   W   W   W   B   W   W    
\arule  W   W   W   B   W   W   W   R   W    
}

\setbox116=\vtop{\leftskip 0pt\parindent 0pt\hsize=\largoo
\baselineskip 10pt

}

\vtop{
\vspace{-15pt}
\begin{tab}\label{rulkill}
\leurre
The rules for the killer .
\end{tab}
\vspace{-5pt}
\ligne{\hfill\box110\hfill\box112\hfill\box114\hfill\box116\hfill}
}
\vskip 5pt
   We remain to look at the change of colour in the signal. 
Rules~236 and~243 applying to the neighbours sharing side~6 and~7 of the central cell respectively
detect the locmotive coming from a fork: either from that of the passive memory switch or from 
that of the flip-flop. Rules~222 and~223 change the colour of the signal: from blue to red
and from red to blue respectively. As mentioned in the implementation, the change to blue when the
signal is red takes two more step that the opposite change.

\vskip 10pt
We arrive to the rules for the fork.
Rule~267 allows the locmotive to arrive at the central cell of the fork which is also the
central cell in the pictures of Figure~\ref{hca73fork}. Again, rule~167 of the passive
memory switch and rule~85 from the crossing allow the creation of a locomotive both
of the neighbours sharing sides~1 and~7 respectively. Next, rules~277 and~281 of 
Table~\ref{rulfork} allow the new locomotives to leave the cells where they were created.
Later, specific motion rules allow the locomotives ot leave the fork~: rule~279, rule~21
attract the locomotive to a neighbour of its previous place and rules~283 and rule~22 allow
it to leave that second place.

\vskip 5pt
\setbox110=\vtop{\leftskip 0pt\parindent 0pt\hsize=\largoo
\baselineskip 10pt
\arule  W   W   R   W   R   R   R   W   W    
\arule  W   W   W   B   R   W   R   R   W    
\arule  W   W   R   W   R   B   W   R   W    
\arule  R   R   W   W   W   W   W   W   R    
\arule  W   W   R   B   W   B   W   R   W    
\arule  R   W   W   W   B   W   W   R   R    
\arule  W   R   W   B   R   B   W   B   R    
\arule  R   W   R   W   W   W   W   R   R    
\ligne{--\hfill}
\arule  W   W   R   B   W   B   R   R   R    
\arule  R   R   W   B   W   B   W   B   W    
\ligne{--\hfill}
\arule  W   W   R   R   R   R   R   W   R    
\arule  R   W   R   B   W   B   W   R   W    
\arule  R   W   R   W   B   W   W   R   R    
\arule  R   W   R   W   R   R   R   W   W    
\arule  R   R   W   R   W   W   B   W   R    
}

\setbox112=\vtop{\leftskip 0pt\parindent 0pt\hsize=\largoo
\baselineskip 10pt
\arule  W   R   R   B   W   B   W   R   W    
\arule  R   R   W   W   B   W   W   R   R    
\arule  R   R   R   W   W   W   W   R   R    
\arule  R   R   R   W   W   W   B   W   R    
\arule  W   R   R   W   R   R   R   R   W    
\arule  R   W   R   B   R   W   R   R   W    
\arule  B   R   R   B   W   W   W   R   B    
\arule  W   R   R   W   R   B   W   R   R    
\arule  R   R   R   W   W   W   W   W   R    
\arule  R   W   R   B   W   B   B   R   W    
\arule  W   W   W   B   R   R   R   R   W    
\arule  R   W   R   W   R   B   W   R   W    
\arule  W   W   R   B   R   B   B   W   W    
\arule  W   W   R   W   R   B   R   R   W    
\arule  R   R   W   W   R   W   W   W   R    
\arule  B   W   R   W   W   W   R   W   B    
}

\vtop{
\vspace{-15pt}
\begin{tab}\label{rulfork}
\leurre
The rules for the fork.
\end{tab}
\vspace{-5pt}
\ligne{\hfill\box110\hskip 20pt\box112\hfill}
}
\vskip 5pt

   With these rules and with this study illustrated bu the figures, we completed the proof
of the following result:

\begin{thm}\label{univ3}
There is a rotation invariant cellular automaton on the heptagrid with $3$~states which is
planar and weakly universal.
\end{thm}

\begin{thebibliography}{5}

\bibitem{fhmmTCS}
F. Herrmann, M. Margenstern,
A universal cellular automaton in the hyperbolic plane,
{\it Theoretical Computer Science},
{\bf 296}, (2003), 327-364.

\bibitem{mmDMTCS}
Cellular Automata and Combinatoric Tilings in Hyperbolic Spaces, a
survey,
{\it Lecture Notes in Computer Sciences}, Editors: Crisitan Calude, Michael J. Dinneen,
V. Vajnovszki, {\bf 2731}, (2003),
48-72,\\ {\tt doi: 10.1007/3-540-45066-1\_4}.

\bibitem{mmbook1}
M. Margenstern,
{\it Cellular Automata in Hyperbolic Spaces, vol. $1$, Theory},
Collection: {\it Advances in Unconventional Computing and Cellular Automata},
Editor: Andrew Adamatzky,
Old City Publishing, Philadelphia, (2007), 422p.

\bibitem{mmbook2}
M. Margenstern,
{\it Cellular Automata in Hyperbolic Spaces, vol. $2$, Implementation and
computations},
Collection: {\it Advances in Unconventional Computing and Cellular Automata},
Editor: Andrew Adamatzky,
Old City Publishing, Philadelphia, (2008), 360p.

\bibitem{mmhepta4}
M. Margenstern,
A universal cellular automaton on the heptagrid of the hyperbolic plane with four states,
{\it Theoretical Computer Science}, {\bf 412}, (2011), 33-56

\bibitem{mmMCUZu}
M. Margenstern,
About Strongly Universal Cellular Automata,
{\it Electronic Proceedings in Theoretical Computer Science}, {\bf 128}(17), (2013), 93-125.

\bibitem{mmbook3}
M. Margenstern,
{\it Small Universal Cellular Automata in Hyperbolic Spaces: A Collection of Jewels},
Collection: {\it Emergence, Complexity and Computation}, Editors: Ivan Zelinka, Andrew Adamtzky,
Guanrong Chen, 
Springer Verlag, (2013), 331p., {\tt doi: 10.1007/978-3-642-36663-5}. 

\bibitem{mmpenta5}
M. Margenstern,
A weakly universal cellular automaton in the pentagrid with five states,
to be published.

\bibitem{mmsyENTCS}
M. Margenstern, Y. Song,
A universal cellular automaton on the ternary heptagrid,
{\it Electronic Notes in Theoretical Computer Science},
{\bf 223}, (2008), 167-185.

\bibitem{mmsyPPL}
A new universal cellular automaton on the pentagrid,
{\it Parallel Processing Letters},
{\bf 19}(2), (2009), 227-246, {\tt doi: 10.1142/S0129626409000195}.

\bibitem{stewart}
I.~Stewart, A Subway Named Turing, Mathematical Recreations in {\it Scientific
American}, (1994), 90-92. 
\end{thebibliography}

\end{document}